\documentclass[11pt,twoside,letterpaper]{article} 

\usepackage{times,fancyhdr}
\usepackage[dvips]{graphicx}

\usepackage{latexsym}
\usepackage{amstext}
\usepackage{amsfonts}
\usepackage{color}
\usepackage{amssymb}
\usepackage{amsmath}
\usepackage{relsize}
\usepackage{amsxtra}
\usepackage{verbatim}
\usepackage{hyperref}
\usepackage{graphicx,subfigure,float}
\usepackage{multirow}

\renewcommand{\thetable}{\arabic{table}}


\makeatletter
\def\cleardoublepage{\clearpage\if@twoside \ifodd\c@page\else%
    \hbox{}%
    \thispagestyle{empty}%
    \newpage%
    \if@twocolumn\hbox{}\newpage\fi\fi\fi}
\makeatother

\def\figurename{Figure}
\makeatletter
\renewcommand{\fnum@figure}[1]{\figurename~\thefigure.}
\makeatother

\def\tablename{Table}
\makeatletter
\renewcommand{\fnum@table}[1]{\tablename~\thetable.}
\makeatother

\begin{document}

\title{
{\begin{flushleft}
\vskip 0.45in
{\normalsize\bfseries\textit{Chapter~}}
\end{flushleft}
\vskip 0.45in
\bfseries\scshape Ground and Applied{-}Field{-}Driven Magnetic States of Antiferromagnets}}

\author{\bfseries\itshape Hai{-}Feng Li$^{1}$ \thanks{E{-}mail address: haifengli@um.edu.mo} and Zikang Tang$^{1}$ \thanks{E{-}mail address: zktang@um.edu.mo} \\
$^{1}$Joint Key Laboratory of the Ministry of Education, \\ Institute of Applied Physics and Materials Engineering,\\ University of Macau, Macao, China}
\date{}
\maketitle
\thispagestyle{empty}
\setcounter{page}{1}

\thispagestyle{fancy}
\fancyhead{}
\fancyhead[L]{In: Book Title \\
Editor: Editor Name
} 
\fancyhead[R]{ISBN: 0000000000  \\
\copyright~2019 Nova Science Publishers, Inc.}
\fancyfoot{}
\renewcommand{\headrulewidth}{0pt}

\begin{abstract}


As discussed in this chapter, we develop a mean{-}field mathematical method to calculate the ground states of antiferromagnets and better understand the applied magnetic{-}field induced exotic properties. Within antiferromagnetic materials competitive and cooperative interactions exist leading to substance extraordinary magnetic states. Our calculations predict that applying a magnetic field to antiferromagnets can switch it from one magnetic state to another. These include antiferromagnetic ground state, spin{-}flop transition, spin{-}flopped state, spin{-}flip transition and spin{-}flipped state. Our framework successfully demonstrates these phase changes. With this, a map of all equilibrium magnetic ground states, as well as the respective equilibrium phase conditions, are derived. Our study provides insight into the origins of the various magnetic states.
\end{abstract}


\maketitle

\pagestyle{fancy}
\fancyhead{}
\fancyhead[EC]{\it Hai{-}Feng Li and Zikang Tang}
\fancyhead[EL,OR]{\thepage}
\fancyhead[OC]{\it Ground and Applied{-}Field{-}Driven Magnetic ...}
\fancyfoot{}
\renewcommand\headrulewidth{0.pt}

\section{Introduction}

Strongly{-}correlated electron materials are given that name due to the fact that a huge amount of electrons within them interact with each other to give the substance nontrivial properties such as the high $T_\textrm{c}$ superconductivity \cite{Bednorz1986, Simon1988, Gao1994, Tanaka2006, Monthoux2007, Paglione2010}, multiferroicity \cite{Schmid1994, Fiebig2002, Kimura2003, Prellier2005, Spaldin2005, Eerenstein2006, Cheong2007, Scott2015}, colossal magnetoresistance effect \cite{Santen1950, Jin1994, Ibarra1995, Kadomtseva2000, Abramovich2001, Demin2002}, metal{-}insulator transition \cite{Andersson1956, Morin1959, Clark1974, Husmann1996, Abrahams1996, Imada1998, Schilbe2002}, or giant magnetostriction effect \cite{Ibarra1995, Abramovich2001, Demin2006}. Most of these extraordinary macroscopic functionalities are causally linked with magnetism. For example, the unconventional superconductivity appears normally in materials where the long{-}range antiferromagnetic (AFM) states are partially or completely suppressed upon hole{-} or electron{-}doping, or by pressure. Magnetism is thus connected with the unconventional superconductivity \cite{Mignod1991, Mook1993, Dai2000, Lake2001, Demler2001, Lake2005, Christianson2008, Lumsden2009, Chi2009, Qiu2009, Haug2009, Parker2010, Inosov2010, Li2010, Zhao2010, Wen2010, Bao2010, Li2011, Scalapino2012, Stock2012, Raymond2012, Eschrig2012, Steffens2013, Zhang2013, Zhang2013-1, Zhang2013-2, Tucker2012, HFLi2011, HFLi2010, Diallo2010}. A full and consistent understanding of these interesting experimental observations has yet to be achieved \cite{Ramirez1997, Rao1998, Kaplan1999, Coey1999, Tokura1999, Tokura1999-1, Tokura1999-2, Okimoto2000, Cooper2001, Dagotto2001, Chatterji2004, Greiner2002, Senthil2004, Coleman2005, Ronnow2005, Zwierlein2006, Sachdev2008, Balakirev2009, Simon2011, Zyuzin2011, chen2015, Kirklin2015, Yang2016, Urban2016} and necessitates a complete reveal of all possible magnetic ground states, and especially magnetic{-}field driven phase transitions and fluctuations of magnets. This is the central topic of this chapter.

Materials consist of atoms. They don{'}t behave as completely free ions in a crystal. On the contrary, they all interact with each other or their surroundings. Such interactions are sometimes large and significant and cannot be overlooked, especially for magnetic ions. Magnetic interactions such as the dipolar, direct exchange, superexchange, double exchange, anisotropic exchange, or the Ruderman{-}Kittel{-}Kasuya{-}Yosida exchange interaction play an important role in bridging the communications between spin moments in a crystal{---}thereby leading to potential long{-}range ordered magnetic states like ferromagnetism or antiferromagnetism.

\begin{figure*}[!t]
\centering \includegraphics[width = 0.88\textwidth] {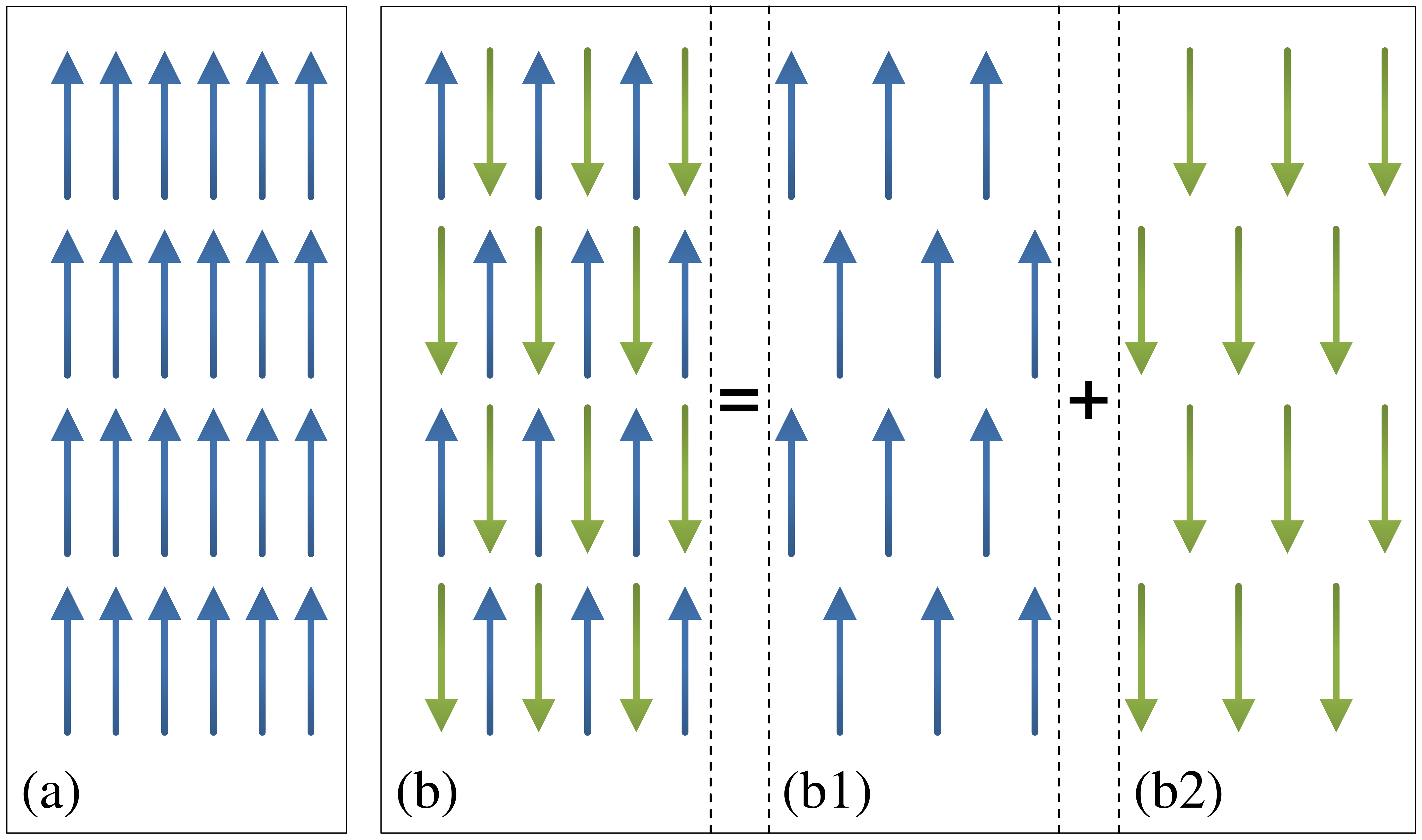}
\caption{
{FM and AFM ground states.}
(\textbf{a}) Spin arrangements in ordered FM state. FM = Ferromagnetic.
(\textbf{b}) Spin arrangements in ordered AFM state. AFM = Antiferromagnetic.
(\textbf{b1 \& b2}) Two interpenetrating FM{-}like sublattices decomposed from (\textbf{b}).
}
\label{Li-Figure1}
\end{figure*}

\begin{figure*}[!t]
\centering \includegraphics[width = 0.88\textwidth] {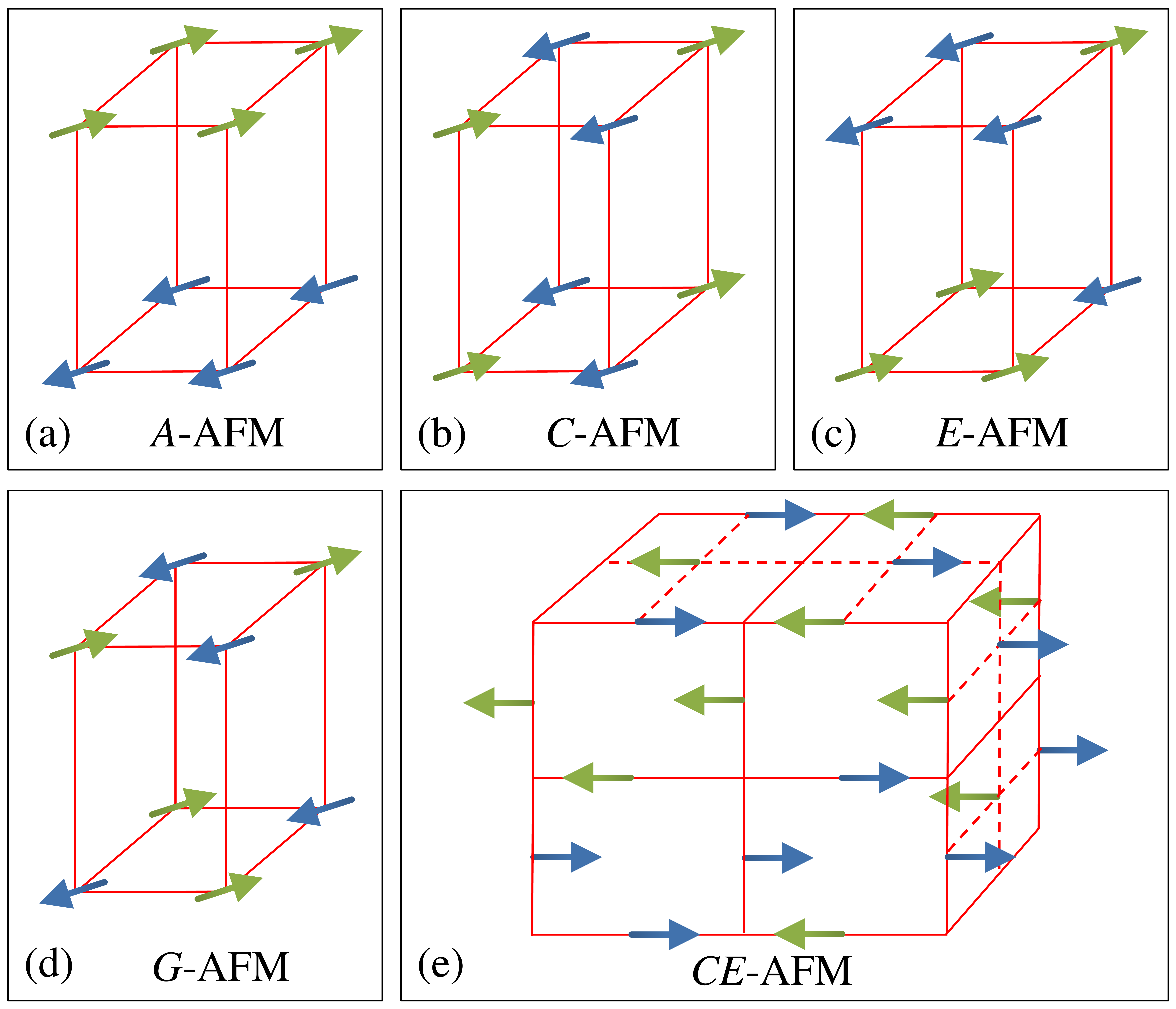}
\caption{
{Possible AFM spin orders. AFM = Antiferromagnetic.}
\emph{A}{-}type (\textbf{a}), \emph{C}{-}type (\textbf{b}), \emph{E}{-}type (\textbf{c}), \emph{G}{-}type (\textbf{d}) and \emph{CE}{-}type (\textbf{e}) AFM spin orders.
The two types of arrows drawn in figures (\textbf{a{-}e}) represent the two possible spin states.
}
\label{Li-Figure2}
\end{figure*}

All spin moments within a ferromagnet, as schematically shown in Figure~\ref{Li-Figure1}(a), are in parallel alignment, producing a spontaneous magnetization even in the absence of an applied field. While in an AFM state, they lie in antiparallel alignment for adjacent spin moments (Figure~\ref{Li-Figure1}(b)). The spin arrangements in an antiferromagnet can be decomposed into two ferromagnetic (FM){-}like interpenetrating states as shown in Figures~\ref{Li-Figure1}(b1) and \ref{Li-Figure1}(b2), where two orientations of the spin moments in sublattices are completely opposite. The strong coupling of spin, charge, orbital and lattice degrees of freedom leads to various long{-}range spin ordered states \cite{Li2008, Xiao2013, Rasmus2012, LSMO2009}. Some of the frequent AFM ground states are schematically illustrated in Figure~\ref{Li-Figure2}. These magnetic states are strongly dependent on the chemical pressure and temperature \cite{LSMO2007-1, LSMO2007-2}. For example, at room temperature, Nd$_{1{-}x}$Sr$_x$MnO$_3$ \cite{Li2008} shows the FM state for 0.3 $<$ x $<$ 0.5. When doped further, the compounds show the \emph{A}{-}type AFM state for 0.5 $<$ x $<$ 0.7. When the doping level is above 0.7, the \emph{C}{-}type AFM state appears \cite{Tokura1999-3, Kajimoto1999}. Above 77 K, YVO$_3$ shows the \emph{C}{-}type AFM spin order, followed by a transition to the \emph{G}{-}type AFM state below 77 K \cite{Blake2002}.

For the traditional spin{-}flop (SFO) transition, typically a first{-}order (FO) type in character, and limited to a collinear antiferromagnet below the N\'{e}el transition temperature, the AFM sublattice spins suddenly rotate 90$^{\circ}$ so that they are perpendicular to the original AFM easy axis while applying a magnetic field (\emph{\textbf{B}}) along the AFM easy axis, and when the applied magnetic-field reaches a critical value $(B_{\textrm{SFO}})$. For the so{-}called spin{-}flip (SFI) transition \cite{Blundell2001, HFLi2016}, as the field strength further increases $(B > B_{\textrm{SFO}})$, the flopped spins gradually tilt toward the field direction until they are completely aligned at a sufficiently high magnetic field $(B_{\textrm{SFI}})$. These magnetic{-}field driven phase transitions are schematically shown in Figure~\ref{Li-Figure3}.

It is hard to experimentally identify the nature of a SFO transition (FO or second{-}order (SO) type) because of the technically unavoidable effect of misalignment between the relevant AFM easy axis and an applied{-}field direction. In 1936, N\'{e}el \cite{Neel1936} theoretically proposed the possibility for a SFO transition. It was then experimentally observed in a CuCl$_2${$\cdot$}2H$_2$O single crystal \cite{Poulis1951}. After that, the SFO phase transition has been extensively investigated, and the phenomenological theory has also been comprehensively developed \cite{Yano1992, Keren1997, Zhang1998, Kurata2000, Sousa2003, Deviatov2004, One2004, Wurtz2005, Witzel2007-1, Witzel2007, Takahashi2008, Cywinski2009, Poggio2009, Yang2009, Phelps2009, Chekhovich2010, Fehr2014, Kim2015, Glaetzle2015, Yokosuk2016, Hou2017, Becker2017, Bender2017, Boldyrev2017, Quiumzadeh2017, Rosenfeld2018, Eve2019}. It is generally confirmed that the SFO phase transition is of FO type \cite{Rasmus2012, Anderson1964, Ranicar1967, Tian2010, Schelleng1969, Shapira1970, Rohrer1975, Butera1981, Filho1991, Bogdanov2007, Holmes1969, Saito1996, Wolf1999, Tsukada2001, Tsukada2003, Zou2015}. On the contrary, some observed SFO transitions \cite{Filho1991, Oliveira1978, Becerra1993} don{'}t show a magnetic hysteresis effect that is the characteristic feature of a FO phase transition. The broad SFO phase transition was attributed to a low magnetic anisotropy \cite{Filho1991, Becerra1993}, a softening of surface magnons \cite{Keffer1973}, a domain effect, or a misalignment of applied magnetic field with regard to the AFM easy axis \cite{Rohrer1975, King1979, Lynn1977}. It is pointed out that such types of continuous magnetic phase transitions \cite{Oh2014, Yokosuk2015}, the absence of magnetic hysteresis, and the experimental observation of a possible intermediate phase in the SFO compound CoBr$_2${$\cdot$}6[0.48D$_2$O, 0.52H$_2$O] \cite{Smeets1982, Basten1980}, cast considerable doubt on the nature of SFO transitions. Actually, it may indicate a SO type phase transition. It is hard to distinguish experimentally the origin of the SO{-}type{-}like SFO transition \cite{Rohrer1975, King1979, Lynn1977} since the allowed small misalignment for a FO SFO transition is usually beyond the present experimental accuracy, and any larger misalignment may change a FO{-}type into a SO{-}type broadening SFO transition. Early theoretical calculations predicted an intermediate regime bordering with the AFM and spin{-}flopped states \cite{Yamashita1972, Liu1973, Becerra1974}. These predications have not yet been confirmed based on the principle of minimum total potential free energy, and most of the theoretical models overlooked the singe{-}ion anisotropy. This is extremely important in the case of lanthanides and actinides \cite{Li2014Tb, Li2015Tm, Wen2015, Li2014Er, LiErPd, Li2012}. Additionally, a rotating FM phase was also predicted for a SFO antiferromagnet \cite{Yamashita1972, Prystasz1982}. However, this kind of unusual magnetic phase has never been observed experimentally, which renders the validity of the phase undecided \cite{HFLi2016}.

Here a mathematical method for a better understanding of the exotic properties of magnetic materials was proposed. We have developed calculations that predict the magnetic{-}field driven SFO and SFI phase transitions of localized collinear antiferromagnets. Our model unifies all possible magnetic ground and excited states and reveals some interesting magnetic phase transitions and coexistences of some of the magnetic states. This study unambiguously reveals a SO{-}type SFO transition. We conclusively rule out the possibility for a rotating FM{-}like magnetic state \cite{Yamashita1972}. This model calculation consistently covers all possible magnetic{-}field driven magnetic states of collinear antiferromagnets, which sheds light on the origins of the various magnetic states. We further deduce an alternative to the estimation of magnetic exchange parameters \cite{HFLi2016}.

\section{Materials and Methods}
\begin{figure*}[!t]
\centering \includegraphics[width = 0.88\textwidth] {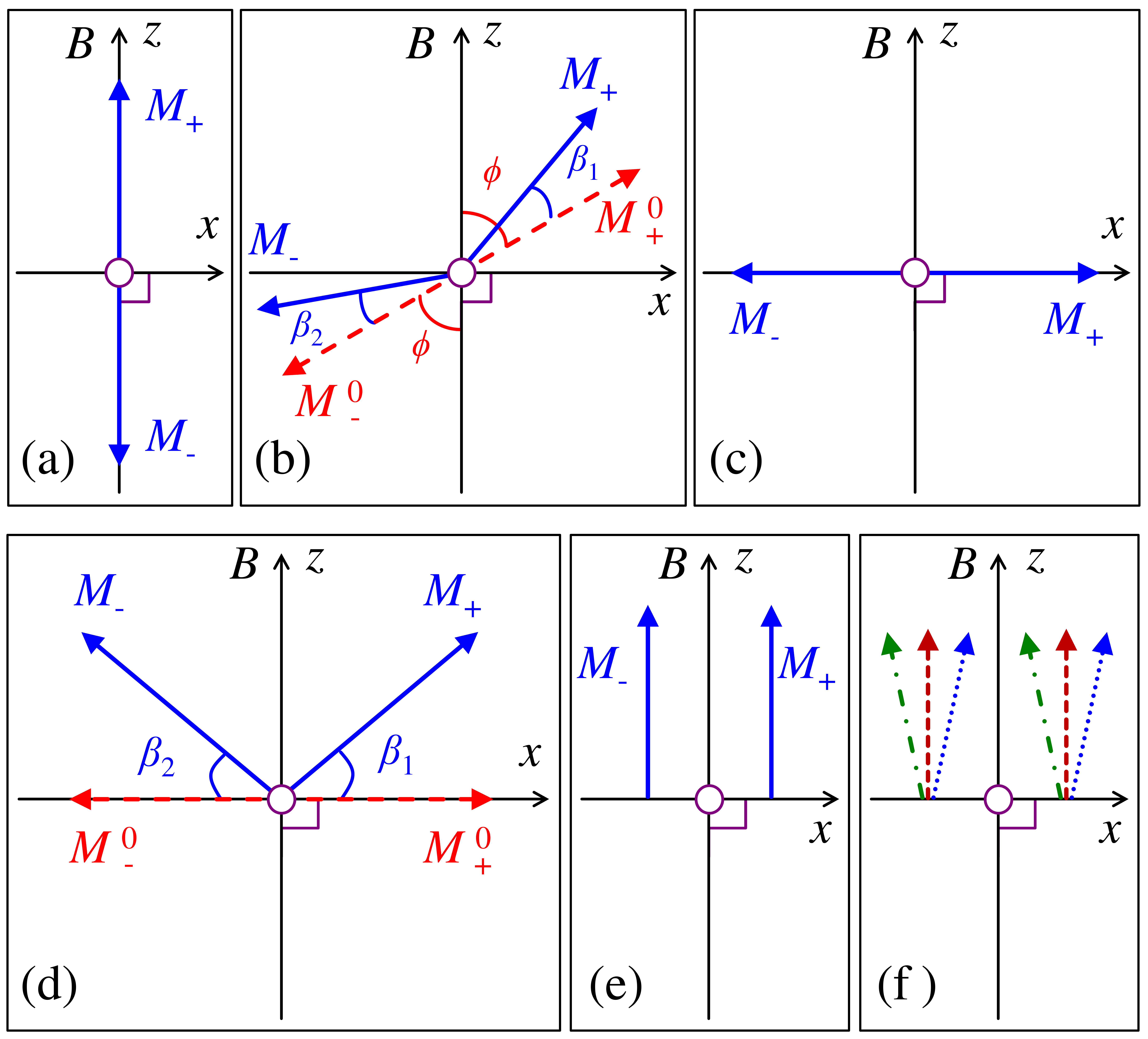}
\caption{
{Magnetic phase transitions while applying magnetic field to a collinear two{-}sublattice antiferromagnet.}
(\textbf{a}) In a normal AFM state $[J > 0$ and $J > \frac{1}{2}(2D {-} \gamma)$], the AFM easy axis $M^0_{-}M^0_{+}$ coincides with the localized sublattice moments $M_{+}$ and $M_{-}$, and all are supposed to be parallel to the $z$ axis when $0 \leq B < B_{\textrm{SFO}}$. Here $\phi = \beta_1 = \beta_2 = 0^{\circ}$.
(\textbf{b}) A SO SFO transition in the range of magnetic fields $B_{\textrm{SFOB}} \leq B < B_{\textrm{SFOF}}$. $\phi$ denotes an angle of the AFM easy direction away from the $z$ axis. $\beta_1$ and $\beta_2$ correspond to the angles of sublattice moments $M_{+}$ and $M_{-}$ away from the $M^0_{-}M^0_{+}$ axis, respectively. Here $0 < \phi < 90^{\circ}$, and $\beta_1 \equiv \beta_2 = \beta$ in the saturation magnetic state at sufficiently low temperatures.
(\textbf{c}) Traditional spin{-}flopped state where $\phi = 90^{\circ}$ and $\beta = 0^{\circ}$.
(\textbf{d}) When $\phi = 90^{\circ}$, the sublattice moments are flopped at $B_{\textrm{SFO}}$ and then tilted away from the $x$ axis by an angle of $\beta$. Here $\beta \in (0^{\circ}, 90^{\circ})$ and $B_{\textrm{SFO}} < B < B_{\textrm{SFI}}$ in the process of a SFI transition.
(\textbf{e}) The sublattice moments $M_{+}$ and $M_{-}$ are completely aligned along the \emph{\textbf{B}} (i.e., $z$) direction in a strong enough magnetic field $B_{\textrm{SFI}}$ so that $\beta = 90^{\circ}$.
It is stressed that in this study, for an antiferromagnet $J > 0$, and the signs ($\pm$) of magnetic exchanges are expressed by the corresponding angles of $\phi$ and $\beta$.
(\textbf{f}) Derived superparamagnetic state where $\beta = 90^\circ$ and $J = 0$.
}
\label{Li-Figure3}
\end{figure*}

We limit our calculation to purely{-}localized collinear AFM systems and overlook the effect of valence electrons on magnetic couplings. In this case, for  two{-}sublattice AFM spins (Figure~\ref{Li-Figure3}), the Hamiltonian terms consist of four parts: magnetic exchange, spin{-}exchange anisotropy, single{-}ion anisotropy and Zeeman coupling to an external magnetic field. We assume that both AFM easy direction $M^0_{-}M^0_{+}$ and localized sublattice moments $M_{+}$ and $M_{-}$ are along the \emph{z} axis (Figure~\ref{Li-Figure3}(a)), and the subsequent completely{-}flopped spins are parallel to the \emph{x} axis (Figures~\ref{Li-Figure3}(c)), thus, the sublattice{-}moment vectors within the \emph{xz} plane (Figure~\ref{Li-Figure3}(b)) can be written as \cite{HFLi2016}:
\setlength\arraycolsep{0.6pt} 
\begin{eqnarray}
\left \{
\begin{array} {ll}
\begin{split}
\label{MVectors}
\widehat{M}_{+} &=  M_{+}[\hat{x} sin(\phi {-} \beta_1) {+} \hat{z} cos(\phi {-} \beta_1)] \textcolor[rgb]{1.00,1.00,1.00}{,} \textcolor[rgb]{1.00,1.00,1.00}{a} \textrm{and}                                                                                                       \\
\widehat{M}_{-} &= {-}M_{-}[\hat{x} sin(\phi {+} \beta_2) {+} \hat{z} cos(\phi {+} \beta_2)],
\end{split}
\end{array}
\right.
\end{eqnarray}
respectively, where $\hat{x}$ and $\hat{z}$ are the unit vectors along the $x$ and $z$ axes, respectively. The angles $\phi$, $\beta_1$ and $\beta_2$ are defined in Figure~\ref{Li-Figure3}. Therefore, the sublattice{-}moment related free energy (\emph{E}) within a mean{-}field approximation can be calculated by \cite{HFLi2016}:
\begin{eqnarray}
\label{energy1}
\begin{split}
\emph{E} = & \textcolor[rgb]{1.00,1.00,1.00}{1}\emph{J} \emph{M}_{+}{\cdot}\emph{M}_{-} {+} \gamma \emph{M}_{+}^z \emph{M}_{-}^z {-} D[(\emph{M}_{+}^z)^2 {+} (\emph{M}_{-}^z)^2] {-} B (\emph{M}_{+}^z {+} \emph{M}_{-}^z)               \\
= & {-}\emph{J} M_{+}M_{-} \cos(\beta_1 {+} \beta_2) {-} \gamma M_{+}M_{-} \cos(\phi {-} \beta_1) \cos(\phi {+} \beta_2)                               \\
& {-} D [M^2_{+} \cos^2(\phi {-} \beta_1) {+} M^2_{-} \cos^2(\phi {+} \beta_2)] \\
&{-} B [M_{+} \cos(\phi {-} \beta_1) {-} M_{-} \cos(\phi {+} \beta_2)],
\end{split}
\end{eqnarray}
where the four terms in turn represent the four Hamiltonian components, and \emph{J} $(> 0)$, $\gamma$ and \emph{D} are the AFM coupling, anisotropic exchange and single{-}ion anisotropic energies, respectively. In an unsaturation magnetic state, with increasing magnetic field \emph{\textbf{B}} $(\parallel z$ axis) as shown in Figures~\ref{Li-Figure3}(a) and \ref{Li-Figure3}(b), the sublattice moment $M_{+}$ ($M_{-}$) increases (decreases) as a consequence, which leads to $\beta_1 < \beta_2$. At the lowest temperature \emph{T} = 0 K, i.e., in a real saturation magnetic state, $M_{+} \equiv M_{-} = M_0$, and thus $\beta_1 \equiv \beta_2 = \beta$. Hence, equation~(\ref{energy1}) can be simplified as \cite{HFLi2016}:
\begin{eqnarray}
\label{energy2}
\begin{split}
\emph{E} = & {-}\emph{J} M_0^2 \cos(2\beta) {-} \gamma M_0^2 \cos(\phi {-} \beta) \cos(\phi {+} \beta) {-} D M_0^2[\cos^2(\phi {-} \beta)      \\
& {+}  \cos^2(\phi {+} \beta)] {-} B M_0[\cos(\phi {-} \beta) {-} \cos(\phi {+} \beta)]                                                        \\
= & {-}\emph{J} M_0^2 \cos(2\beta) {-} \frac{\gamma M_0^2}{2} [\cos(2\phi) {+} \cos(2\beta)] {-} D M_0^2[1                                     \\
& {+} \cos(2\phi) \cos(2\beta)] {-} 2 B M_0 \sin \phi \sin \beta.
\end{split}
\end{eqnarray}

\section{Results}
\subsection{Equilibrium Magnetic State}
Possible magnetic ground and excited states can be derived from different combinations of the FO partial differential equations, i.e., $\frac{\partial E}{\partial \beta} = \frac{\partial E}{\partial \phi} =$ 0 with equation~(\ref{energy2}) \cite{HFLi2016}:
\setlength\arraycolsep{0.6pt} 
\begin{eqnarray}
\label{AFM-SFO}
\left.
\begin{array} {r r}
2 D M_0 \sin \phi \cos (2 \beta) + \gamma M_0 \sin \phi - B \sin \beta = 0                                                                    \\
2 J M_0 \sin \beta + 2 D M_0 \cos (2 \phi) \sin \beta + \gamma M_0 \sin \beta - B \sin \phi = 0
\end{array}
\right\};                                                                                                                                        \\
\label{SFO}
\left.
\begin{array} {r r}
\cos \phi = 0                                                                                                                                 \\
2 J M_0 \sin \beta + 2 D M_0 \cos (2 \phi) \sin \beta + \gamma M_0 \sin \beta - B \sin \phi = 0
\end{array}
\right\};                                                                                                                                        \\
\label{FM}
\left.
\begin{array} {r r}
2 D M_0 \sin \phi \cos (2 \beta) + \gamma M_0 \sin \phi - B \sin \beta = 0                                                                    \\
\cos \beta = 0
\end{array}
\right\};                                                                                                                                        \\
\label{SFI}
\left.
\begin{array} {r r}
\cos \phi = 0                                                                                                                                  \\
\cos \beta = 0
\end{array}
\right\}.
\end{eqnarray}

We solved the four combinations~(\ref{AFM-SFO}{-}\ref{SFI}) and correlated the results with physical meanings accordingly \cite{HFLi2016}.

(i) From the combination~(\ref{AFM-SFO}), we obtained two solutions:
\begin{eqnarray}
\label{A}
\left.
\begin{array} {r r}
\textrm{(A)} \sin \phi = \sin \beta = 0, i.e., \phi = \beta = 0;           \\
\label{B}  \\
\textrm{(B)} \sin \phi = \delta \sin \beta, \textrm{where } \sin \beta = \sqrt{\frac{\delta M_0 (2D {+} \gamma) {-} B}{4 \delta M_0 D}}\textcolor[rgb]{1.00,1.00,1.00}{,}    \\
\\
\textrm{and } \delta = \sqrt{\frac{2 J {+} 2 D {+} \gamma}{2D {+} \gamma}}. \\
\end{array}
\right\}
\end{eqnarray}

Here the case (A) is related to an AFM ground state (Figure~\ref{Li-Figure3}(a)). The case (B) indicates a correlated change of $\phi$ with $\beta$. In Figure~\ref{Li-Figure3}, $0^{\circ} \leq \phi \leq 90^{\circ}$. As a result, there exist two boundary magnetic fields that correspond to the second solution of the combination~(\ref{AFM-SFO}). When $\phi = 0$, $\sin \phi = \delta \sin \beta = 0$. One can thus deduce that the initial magnetic field for the beginning of the SFO transition
\begin{eqnarray}
\label{SFOB1}
B_{\textrm{SFOB}} = M_0 \sqrt{(2D {+} \gamma)(2J {+} 2D {+} \gamma)}.
\end{eqnarray}
When $\phi = \frac{\pi}{2}$, $\delta \sin \beta = 1$, the final magnetic field for the ending of the SFO transition
\begin{eqnarray}
\label{SFOF1}
B_{\textrm{SFOF}} = M_0 (2J {-} 2D {+} \gamma) \sqrt{\frac{2D {+} \gamma}{2J {+} 2D {+} \gamma}}.
\end{eqnarray}
When $B_{\textrm{SFOB}} \geq B_{\textrm{SFOF}}$, we can derive the precondition of a FO SFO transition: $D \geq 0$ and $2D {+} \gamma > 0$. Further, when $B_{\textrm{SFOB}} < B_{\textrm{SFOF}}$, i.e., ${-}\frac{1}{2} \gamma < D < 0$, a SO SFO transition surprisingly occurs spontaneously. This originates from a negative single{-}ion anisotropy that is additionally restricted to a certain range by the anisotropic exchange interaction ($\gamma$).

(ii) The combination~(\ref{SFO}) implies that
\begin{eqnarray}
\label{}
\phi &=& \frac{\pi}{2}, \textrm{and } \sin \beta = \frac{B}{M_0(2J {-} 2D {+} \gamma)},
\end{eqnarray}
which corresponds to the process of a SFI transition (Figure~\ref{Li-Figure3}(d)). When
\begin{eqnarray}
\label{SFI1}
B = B_{\textrm{SFI}} = M_0(2J {-} 2D {+} \gamma),
\end{eqnarray}
$\beta = \frac{\pi}{2}$, implying a spin{-}flipped (SFID) state (Figure~\ref{Li-Figure3}(e)). Therefore, the SFI transition field $B_{\textrm{SFI}}$ depends not only on the moment size $M_0$ but also on the values of $J$, $\gamma$ and $D$.

(iii) From the combination~(\ref{FM}), we can deduce that
\begin{eqnarray}
\label{}
\beta &=& \frac{\pi}{2}, \textrm{and } \sin \phi = \frac{B}{M_0(\gamma {-} 2D)}.
\end{eqnarray}
When $\beta = \frac{\pi}{2}$, both sublattice moments $M_{+}$ and $M_{-}$ are perpendicular to the AFM axis $M^0_{-}M^0_{+}$, forming a rotating FM{-}like state. The value of $\phi$ can intrinsically be modified by a change in magnetic field $B$.

(iv) The combination~(\ref{SFI}) indicates $\phi = \beta = \frac{\pi}{2}$. This corresponds to a SFID state (Figure~\ref{Li-Figure3}(e)).

\subsection{Free Energy Calculation}
In the following, we calculated free-energy scales of the derived magnetic states by substituting their respective equilibrium phase conditions back into equation~(\ref{energy2}). We thus obtain \cite{HFLi2016}:
\begin{eqnarray}
\left\{
\begin{split}
\label{AFME}
E_{\textrm{AFM} (z{-}\textrm{axis})} = \text{\textcolor[rgb]{1.00,1.00,1.00}{a}} & {-}(J {+} \gamma {+} 2 D) M_0^2  \text{\textcolor[rgb]{1.00,1.00,1.00}{aa}}  \qquad\qquad (0 \leq B < B_{\textrm{SFO}});   \\
\label{FOSFOE}
E_{x{-}\textrm{axis}} = \text{\textcolor[rgb]{1.00,1.00,1.00}{a}} & {-}J M_0^2                                      \text{\textcolor[rgb]{1.00,1.00,1.00}{1}}   \qquad\qquad\qquad\qquad (\phi = 90^\circ, \beta = 0^\circ);                                                                                                                                                                                                     \\
\label{SFIE}
E_{\textrm{SFI}}  = \text{\textcolor[rgb]{1.00,1.00,1.00}{a}} & {-}J M_0^2 {-} \frac{B^2}{2J {-} 2D {+} \gamma}     \qquad (B_{\textrm{SFO}} \leq B < B_{\textrm{SFI}});                                      \\
\label{SFIDE}
E_{\textrm{SFID}} = \text{\textcolor[rgb]{1.00,1.00,1.00}{a}} & [{-}J {-}(2J {-} 2 D {+} \gamma)]M_0^2              \text{\textcolor[rgb]{1.00,1.00,1.00}{a}} \qquad\qquad (B = B_{\textrm{SFI}});            \\
\label{FMlikeE}
E_{\textrm{FM{-}like}} = \text{\textcolor[rgb]{1.00,1.00,1.00}{a}} & J M_0^2 {-} \frac{B^2}{\gamma {-} 2D}          \qquad\qquad [0 \leq B \leq M_0 (\gamma {-}2 D)];
\end{split}
\right.
\end{eqnarray}
in addition, the one corresponding to the SO SFO transition is:
\begin{eqnarray}
\label{SOSFOE}
\left\{
\begin{split}
E_{\textrm{SO{-}SFO}} = \text{\textcolor[rgb]{1.00,1.00,1.00}{a}} & {-}JM_0^2(1 {-} 2\sin^2\beta) {-} \gamma M^2_0 (1 {-} \sin^2\beta {-} \delta^2\sin^2\beta)                                          \\
\text{\textcolor[rgb]{1.00,1.00,1.00}{a}} & {-} DM^2_0 [1 {+} (1 {-} 2\delta^2\sin^2\beta)(1 {-} 2\sin^2\beta)] {-} 2BM_0\delta\sin^2\beta                                                              \\
\\
& \text{\textcolor[rgb]{1.00,1.00,1.00}{a}}(\sin \beta = \sqrt{\frac{\delta M_0 (2D {+} \gamma) {-} B}{4 \delta M_0 D}}, \delta = \sqrt{\frac{2 J {+} 2 D {+} \gamma}{2D {+} \gamma}};                  \\
& \qquad\qquad\text{\textcolor[rgb]{1.00,1.00,1.00}{aa}} {-}\frac{1}{2}\gamma < D < 0;  B_{\textrm{SFOB}} \leq B < B_{\textrm{SFOF}}).
\end{split}
\right.
\end{eqnarray}

We quantitatively compared the free energies.

\begin{figure*}[!ht]
\centering \includegraphics[width = 0.72\textwidth] {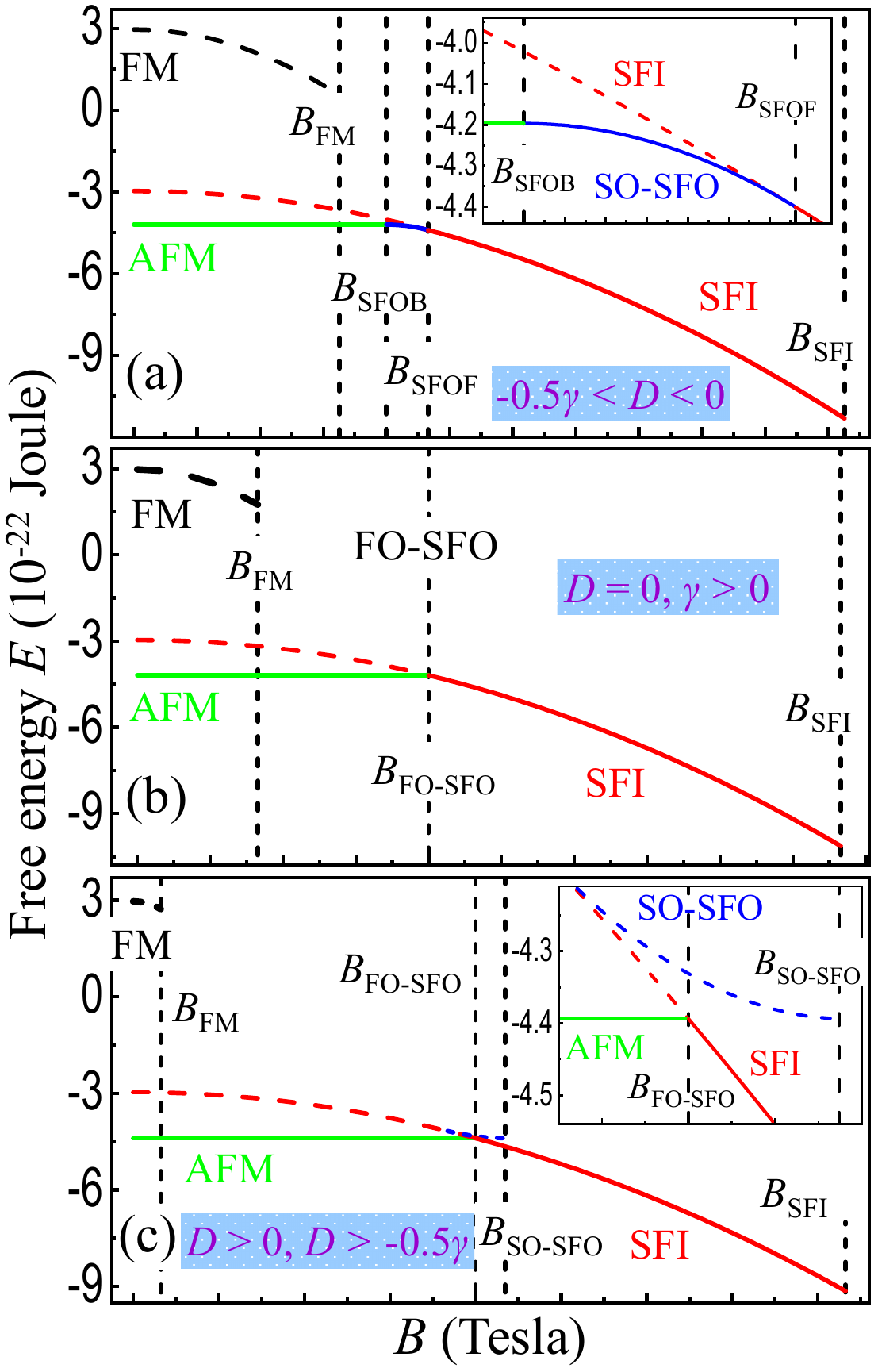}
\caption{{Calculated relative sublattice{-}moment related free energies of the deduced magnetic states as a function of magnetic field \emph{B}.}
(\textbf{a}) When ${-}\frac{1}{2}\gamma < D < 0$, a SO SFO transition occurs. Here we suppose that $M_\textrm{0}$ = 4 $\mu_\textrm{B}$, \emph{J} = 2 T/$\mu_\textrm{B}$, \emph{D} = {-}0.2 T/$\mu_\textrm{B}$, and $B_{\textrm{SFOB}} = 8$ T.
(\textbf{b}) When $D = 0$ and $\gamma > 0$, a FO SFO transition happens. Here we suppose that $M_\textrm{0}$ = 4 $\mu_\textrm{B}$, \emph{J} = 2 T/$\mu_\textrm{B}$, \emph{D} = 0 T/$\mu_\textrm{B}$, and $B_{\textrm{SFOB}} = 8$ T.
(\textbf{c}) When $D > 0$ and $D > {-}\frac{1}{2}\gamma$, a FO SFO transition occurs. Here we suppose that $M_\textrm{0}$ = 4 $\mu_\textrm{B}$, \emph{J} = 2 T/$\mu_\textrm{B}$, \emph{D} = 0.2 T/$\mu_\textrm{B}$, and $B_{\textrm{SFOB}} = 8$ T.
In (\textbf{a}{-}\textbf{c}), the calculated free energies are all plotted for a clear comparison.
The insets of \textbf{a} and \textbf{c} show an enlargement of the most interesting magnetic{-}field regimes. The $E_{\textrm{SFI}}$ under $B < B_{\textrm{SFOF}}$ is also shown (dashed read line, as marked).
In (\textbf{c}), the mathematically permissible \emph{E} of the SO SFO transition is also displayed. In any case, the solid lines as shown in (\textbf{a}{-}\textbf{c}) represent the theoretically{-}allowed magnetic ground states and associated magnetic phase transitions with a change in the strength of magnetic field \emph{\textbf{B}}.
}
\label{Li-Figure4}
\end{figure*}

(i) The case of the SO SFO transition (${-}\frac{1}{2}\gamma < D <0$). Supposing that $M_\textrm{0}$ = 4 $\mu_\textrm{B}$, \emph{J} = 2 T/$\mu_\textrm{B}$, \emph{D} = {-}0.2 T/$\mu_\textrm{B}$, and $B_{\textrm{SFOB}} = 8$ T \cite{Rasmus2012, Peters2009}, which are all substituted into equation~(\ref{SFOB1}), one thus gets $\gamma \sim$ 1.228 T/$\mu_\textrm{B}$ which satisfies the boundary condition $D > {-}\frac{1}{2}\gamma$. Therefore, based on these values, we obtain that $B_{\textrm{SFOF}} \sim$ 9.325 T [equation~(\ref{SFOF1})], $\delta \sim 2.414$ [equation~(\ref{B})], and $B_{\textrm{SFI}} \sim 22.514$ T [equation~(\ref{SFI1})]. Hence, the relative sublattice{-}moment related free energies of all possible magnetic states can be calculated as shown in Figure~\ref{Li-Figure4}(a).

(ii) With the above assumed parameters if one sets $D = 0$ T/$\mu_\textrm{B}$, then $\gamma \sim$ 0.828 T/$\mu_\textrm{B}$, and $B_{\textrm{SFOF}} = B_{\textrm{SFOB}} = 8$ T. This corresponds to a FO SFO transition. The calculated relative free energies at $D = 0$ T/$\mu_\textrm{B}$ are shown in Figure~\ref{Li-Figure4}(b).

(iii) In the case of $D > 0$, $B_{\textrm{SFOB}} > B_{\textrm{SFOF}}$. To extract the exact field for the FO SFO transition, we solved $E_{\textrm{AFM}(z{-}\textrm{axis})}$ = $E_{\textrm{SFI}}$ [equation~(\ref{SFIE})]. This yields
\begin{eqnarray}
\label{SFO1}
B_{\textrm{SFO}} = M_0\sqrt{(2D {+} \gamma)(2J {-} 2D {+} \gamma)}.
\end{eqnarray}
If one sets $D = 0.2$ T/$\mu_\textrm{B}$, then $\gamma \sim$ 0.561 T/$\mu_\textrm{B}$, and $B_{\textrm{SFI}} \sim 16.645$ T. The corresponding free-energy scales at $D = 0.2$ T/$\mu_\textrm{B}$ are displayed in Figure~\ref{Li-Figure4}(c).

\subsection{Nature of the SFO and SFI Transitions}
As shown in Figure~\ref{Li-Figure4}(a), an AFM state persists up to $B_{\textrm{SFOB}}$, then a SO SFO transition occurs in the range of magnetic fields $B_{\textrm{SFOB}} \leq B \leq B_{\textrm{SFOF}}$, followed by a SFI transition at $B > B_{\textrm{SFOF}}$. Finally, all sublattice spins are aligned along the magnetic field direction at $B_{\textrm{SFI}}$. By contrast, as shown in Figures~\ref{Li-Figure4}(b) and \ref{Li-Figure4}(c), an antiferromagnet experiences a FO SFO transition at $B_{\textrm{FO{-}SFO}}$ and then enters directly into the process of a SFI transition. It is pointed out that an occurrence of the SFO transition is attributed to the existence of magnetic anisotropy, $\gamma$ and/or $D$. In the SFOD state \cite{HFLi2016},
\begin{eqnarray}
\label{}
\textrm{sin} \beta = \frac{B_{\textrm{SFO}} (\textrm{or }B_{\textrm{SFOF}})}{M_0 (2J {-} 2D {+} \gamma)}.
\end{eqnarray}
Therefore, the angle $\beta$ can never be zero, which is a sharp contrast to the traditional FO-type SFO transition where $\beta = 90^\circ$ in the SFOD state.

\begin{figure*}[!ht]
\centering \includegraphics[width = 0.88\textwidth] {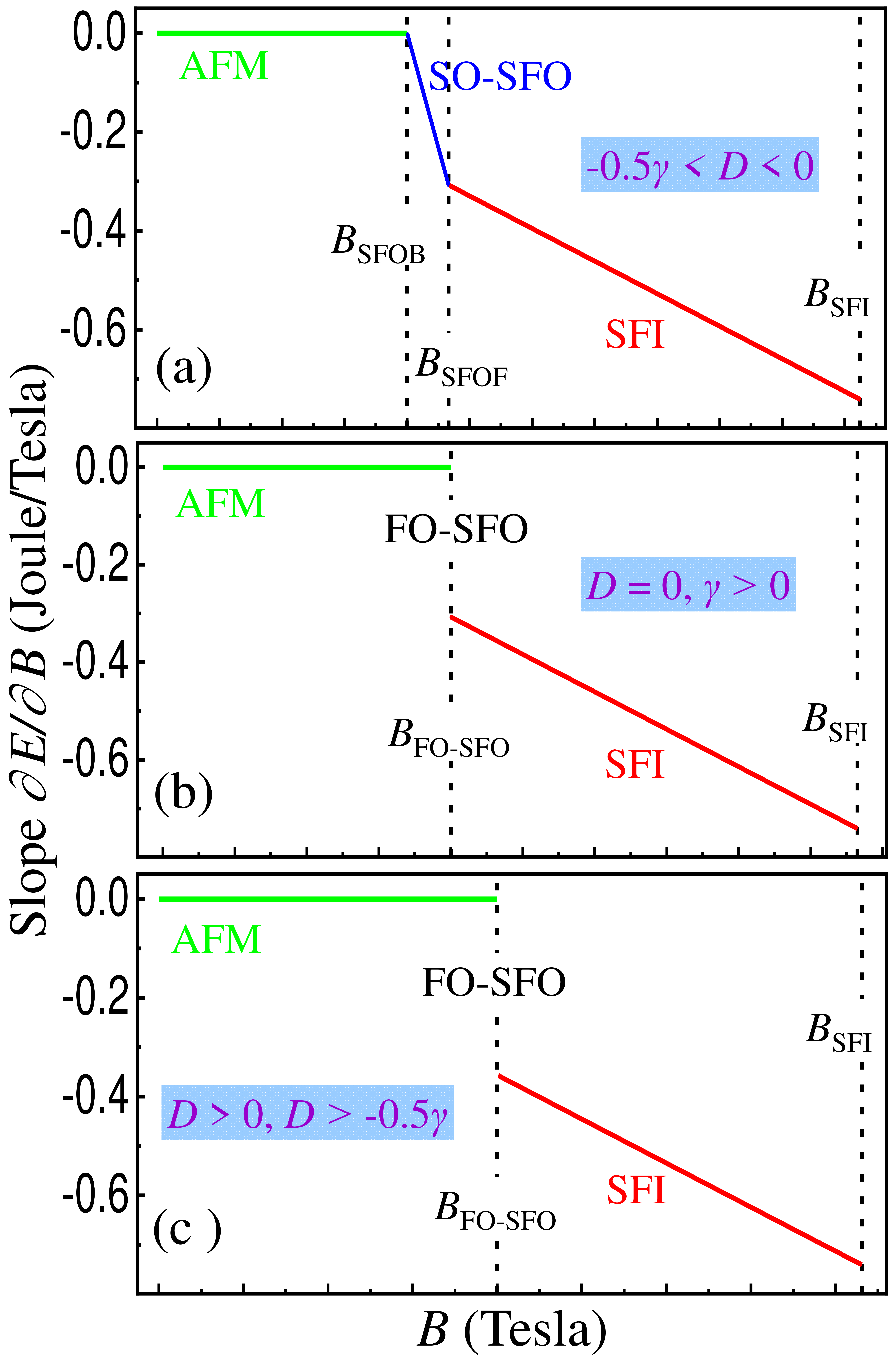}
\caption{{Calculated relative sublattice{-}moment related free-energy slopes of the deduced magnetic states, $\partial E / \partial B$, as a function of magnetic field \emph{B}.}
(\textbf{a}) When ${-}\frac{1}{2}\gamma < D < 0$, the first derivative of the sublattice{-}moment related free energy \emph{E} with regard to magnetic field \emph{B} equals to zero in the AFM state and displays a continuous change while undergoing the SO SFO transition from $B_{\textrm{SFOB}}$ to $B_{\textrm{SFOF}}$, and then the SFI transition from $B_{\textrm{SFO}}$ to $B_{\textrm{SFI}}$.
(\textbf{b}) When $D = 0$ and $\gamma > 0$, and (\textbf{c}) $D > 0$ and $D > {-}\frac{1}{2}\gamma$, an abrupt change in the slope occurs at the FO SFO transition field $B_{\textrm{FO{-}SFO}}$ = $B_{\textrm{SFO}}$.
}
\label{Li-Figure5}
\end{figure*}

\begin{figure*}[!ht]
\centering \includegraphics[width = 0.88\textwidth] {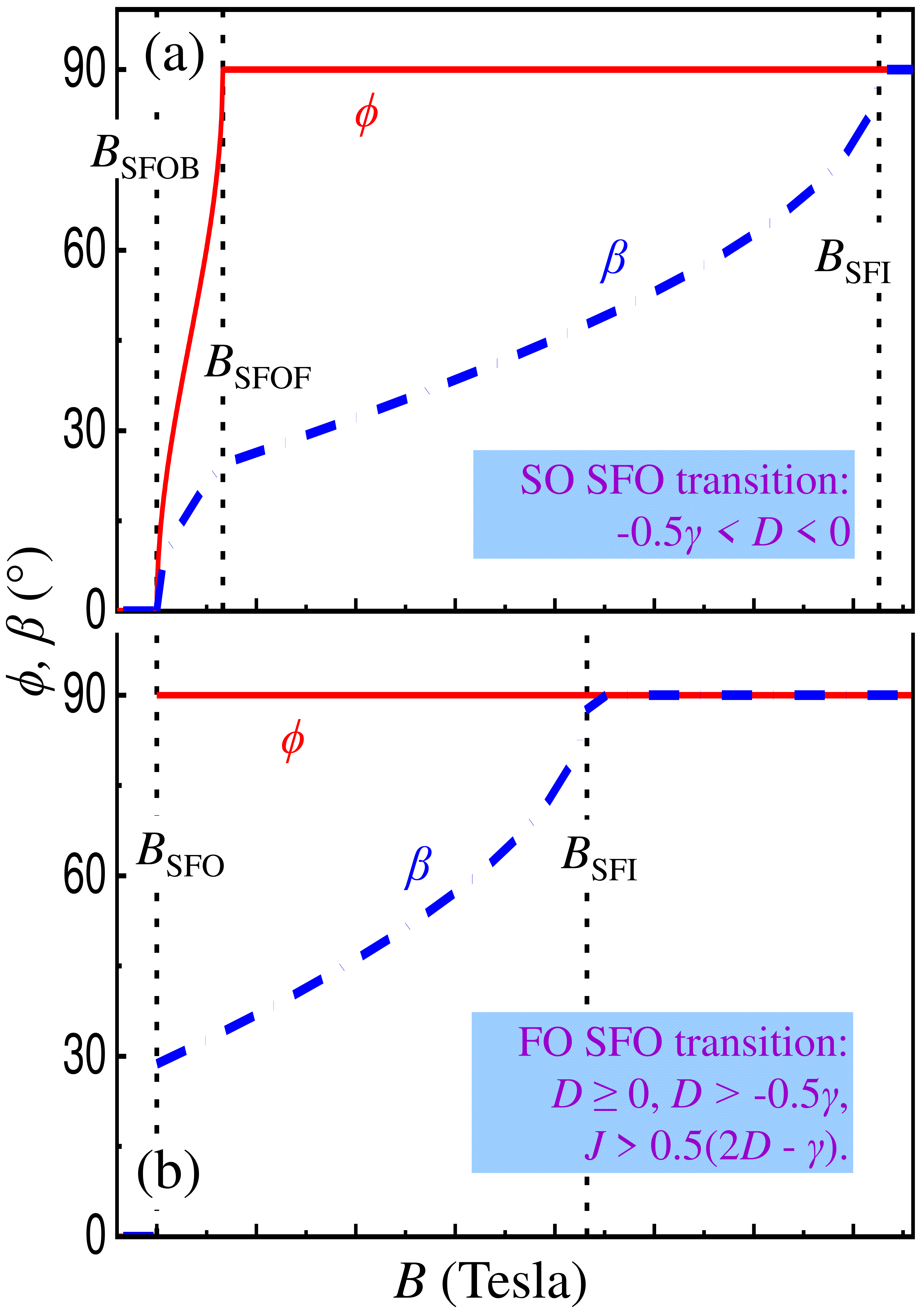}
\caption{
{Variations of the angles $\phi$ and $\beta$ with magnetic filed \emph{B} for SFO transitions.}
(\textbf{a}) When ${-}\frac{1}{2}\gamma < D < 0$, $\phi$ and $\beta$ increase continuously in the range of magnetic fields $B_{\textrm{SFOB}} \leq B < B_{\textrm{SFOF}}$, suggesting a SO SFO transition. When $B_{\textrm{SFO}} \leq B \leq B_{\textrm{SFI}}$, $\phi = 90^{\circ}$, and $\beta$ keeps a continuous increase up to 90$^{\circ}$ at $B_{\textrm{SFI}}$.
(\textbf{b}) When $D \geq 0$ and $D > {-}\frac{1}{2}\gamma$, $\phi$ suddenly increases up to 90$^{\circ}$ at $B_{\textrm{SFOB}} = B_{\textrm{SFOF}} = B_{\textrm{SFO}}$ indicative of a FO SFO transition, after which the magnetic phase enters into the process of a SFI transition.
In \textbf{a} and \textbf{b}, below $B_{\textrm{SFOB}}$ ($B_{\textrm{SFO}}$), $\phi = \beta = 0^{\circ}$; above $B_{\textrm{SFI}}$, $\phi = \beta = 90^{\circ}$.
}
\label{Li-Figure6}
\end{figure*}

We calculated the angles $\phi$ and $\beta$ and further confirmd the FO and the SO SFO transitions. The nature of a SFO transition can also be recognized by the character, continuous or discontinuous, of the first derivative of the free energy (Figure~\ref{Li-Figure5}) with regard to magnetic field based on the Ehrenfest{'}s criterion \cite{Tari2003} for the FO and the SO phase transitions. A continuous slope change is clearly illustrated in Figure~\ref{Li-Figure5}(a), where one can easily deduce that the second derivative $\partial ^2 E / \partial ^2 B$ is indeed discontinuous. By contrast, an abrupt change in the slope is obviously displayed at $B_{\textrm{FO{-}SFO}}$ in Figures~\ref{Li-Figure5}(b) and \ref{Li-Figure5}(c). To better understand the magnetic phase transitions with field, the values of the angles $\phi$ and $\beta$ (Figure~\ref{Li-Figure3}) for all deduced magnetic states are calculated in the whole magnetic field range as shown in Figures~\ref{Li-Figure6}(a) and \ref{Li-Figure6}(b). The SO (Figure~\ref{Li-Figure6}(a)) and the FO (Figure~\ref{Li-Figure6}(b)) SFO transitions are much clear in terms of the variations of $\phi$ and $\beta$ with magnetic field. Finally, it can convincingly be concluded that a SO SFO transition indeed exists theoretically \cite{HFLi2016}.

\section{Discussion}
\subsection{Ruling Out the Rotating FM{-}Like State \cite{HFLi2016}}

We first rule out the rotating FM{-}like state. It is clear that in the magnetic{-}field range $B \leq B_{\textrm{FM}}$, the relative sublattice{-}moment related free energy $E_{\textrm{FM{-}like}}$ is always higher than those of other allowed magnetic states (Figure~\ref{Li-Figure4}), indicating that the rotating FM{-}like state does not exist at all in view of its relatively higher free energy.

\subsection{Equilibrium Phase Condition and Origin of the Magnetic Phase Transition \cite{HFLi2016}}
To clearly present the deduced magnetic ground states and associated magnetic phase transitions with magnetic field, we calculate the three{-}dimensional $(J, \gamma, D)$ and the two{-}dimensional $(\gamma, D)$ phase diagrams as shown in Figures~\ref{Li-Figure8}-\ref{Li-Figure13} and \ref{Li-Figure14}, respectively. The corresponding spin configurations in point are schematically exhibited in Figure~\ref{Li-Figure15}.

In this study, for an antiferromagnet $J > 0$; when $J < 0$, on the other hand, the magnet houses a FM state. Additionally, for the existences of the SFO (FO or SO) and SFI transitions, $B_{\textrm{SFOF}} > 0$ [Figure~\ref{Li-Figure3}(b) and equation~(\ref{SFOB})], $B_{\textrm{SFO}} > 0$ [Figure~\ref{Li-Figure3}(d) and equation~(\ref{SFO8})], and $B_{\textrm{SFI}} > 0$ [Figure~\ref{Li-Figure3}(e) and equations~(\ref{SFOB}) and (\ref{SFO8})]. One thus can deduce that $J > \frac{1}{2}(2D {-} \gamma)$ for the validation of these magnetic states. Furthermore, by comparing $E_{\textrm{AFM}(z{-}\textrm{axis})}$ and $E_{\textrm{SFID}}$ [equation~(\ref{SFIE})], one can finally conclude that there exists the possibility for a FM state even when $J > 0$ as shown in Figures~\ref{Li-Figure13} and \ref{Li-Figure15}(f), where $0 < J < \frac{1}{2}(2D {-} \gamma)$.

From the foregoing remarks, we know that for a FO SFO transition $D \geq 0$ and $D > {-}\frac{1}{2}\gamma$. By including the condition of $J > \frac{1}{2}(2D {-} \gamma)$ for the validated existence of an antiferromagnet, one can divide the FO SFO transition into two regimes:

(i) FO SFO transition{-}1: $D > \pm\frac{1}{2}\gamma$ and $J > \frac{1}{2}(2D {-} \gamma)$ (Figure~\ref{Li-Figure8});

(ii) FO SFO transition{-}2: $0 \leq D \leq \frac{1}{2}\gamma$ (Figure~\ref{Li-Figure9}).

\begin{figure*}[!ht]
\centering \includegraphics[width = 0.88\textwidth] {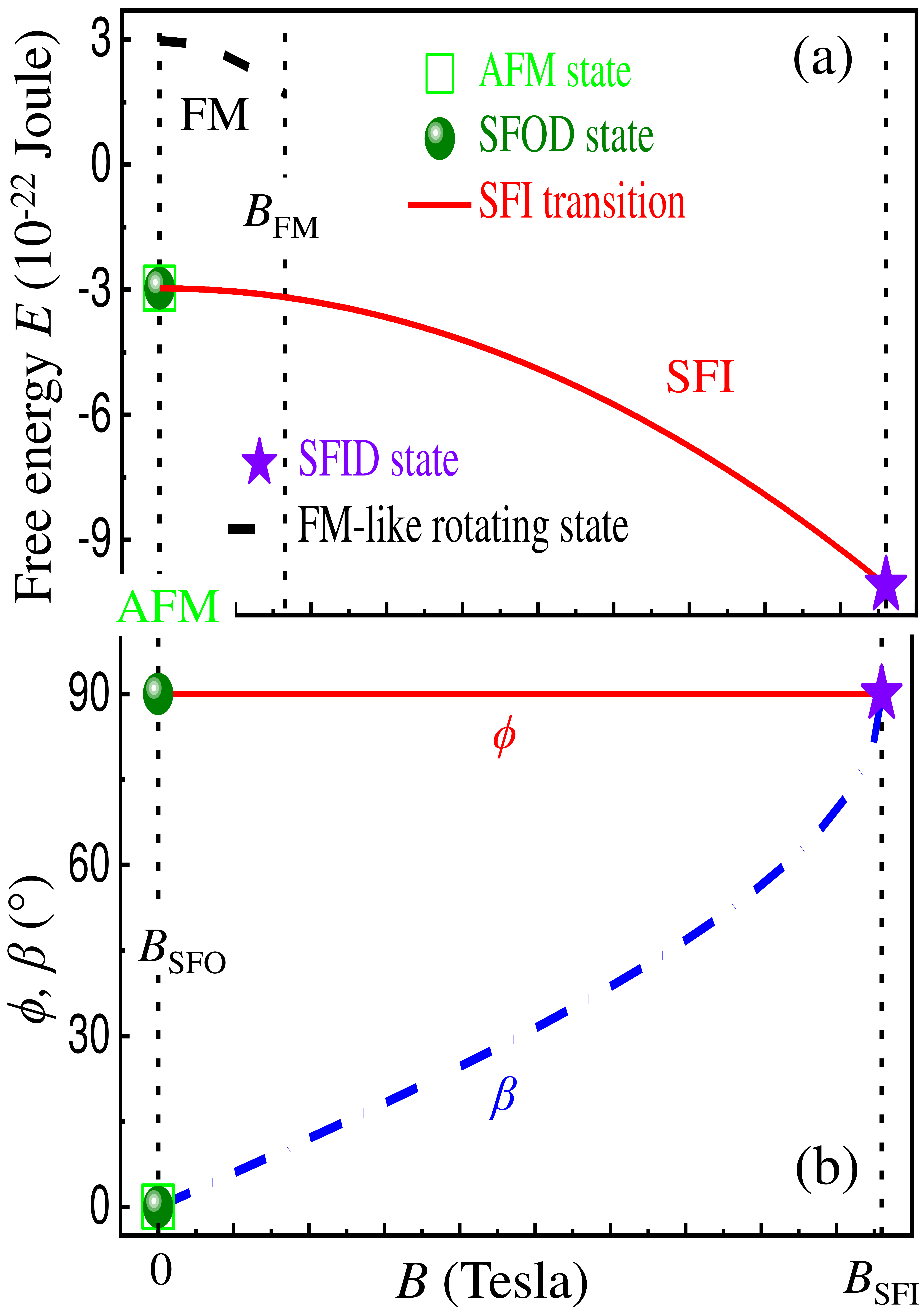}
\caption{
{Variations of the angles $\phi$ and $\beta$ with magnetic filed \emph{B} for a direct SFI transition, corresponding to Figure~\ref{Li-Figure11}.}
(\textbf{a}) The calculated relative sublattice{-}moment related free energies when $D = {-}\frac{1}{2}\gamma$ and $J > {-}\gamma$. In this case, $B_{\textrm{SFO}} = 0$, and thus a SFI transition occurs directly while $B > 0$.
(\textbf{b}) The corresponding variations of the angles $\phi$ and $\beta$ with magnetic field \emph{B}. In \textbf{a} and \textbf{b}, $M_0 =$ 4 $\mu_\textrm{B}$, $J$ = 2 T/$\mu_\textrm{B}$, $D$ = {-}0.2 T/$\mu_\textrm{B}$, and $\gamma = 0.4$ T/$\mu_\textrm{B}$.
}
\label{Li-Figure7}
\end{figure*}

\begin{figure*}[!ht]
\centering \includegraphics[width = 0.88\textwidth] {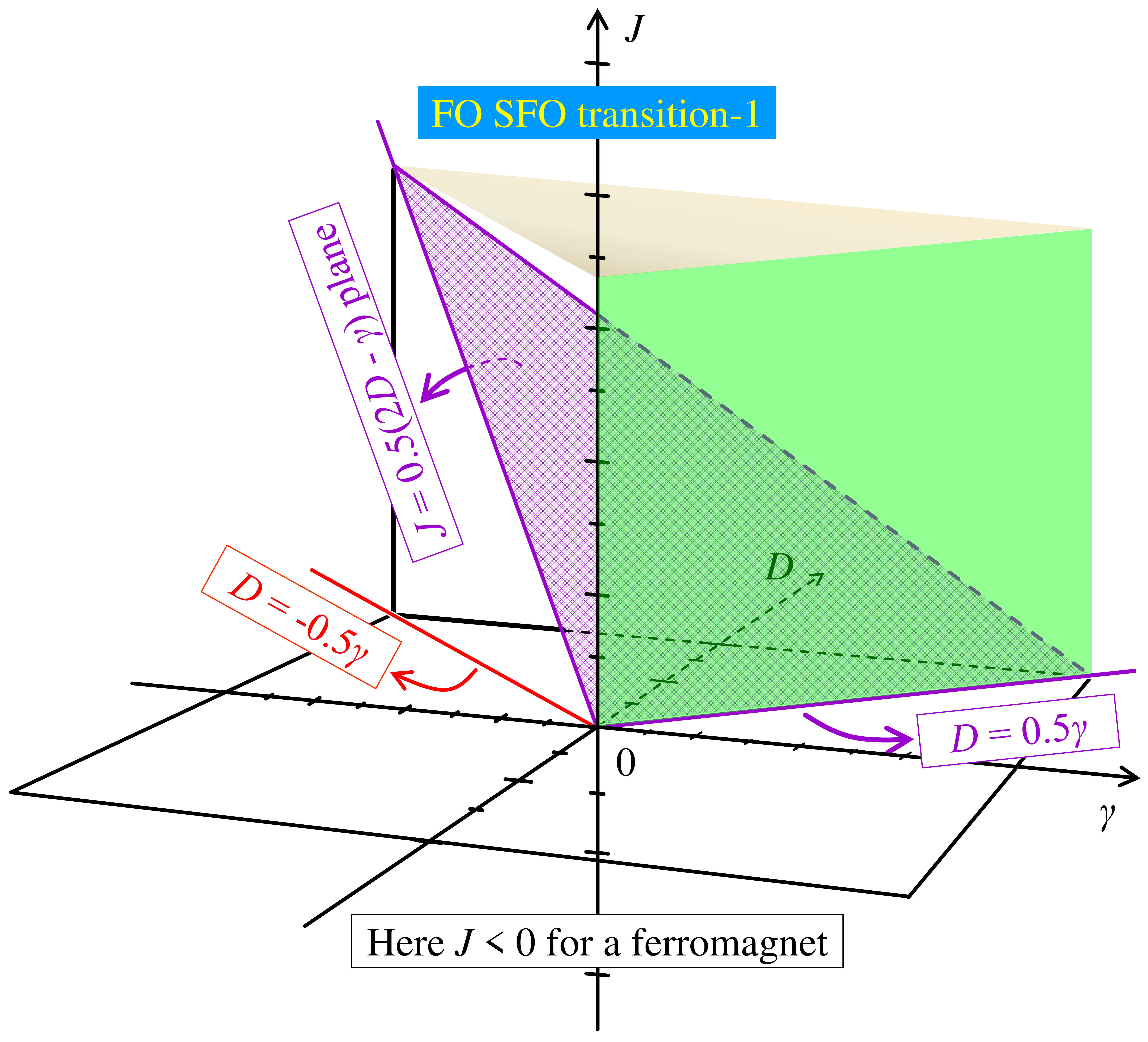}
\caption{
{Three{-}dimensional ($J, \gamma, D$) phase diagram of a collinear two{-}sublattice antiferromagnet.} When $D > \pm 0.5\gamma$ and $J > \frac{1}{2}(2D {-} \gamma)$, a FO SFO transition{-}1 occurs.
}
\label{Li-Figure8}
\end{figure*}

\begin{figure*}[!ht]
\centering \includegraphics[width = 0.88\textwidth] {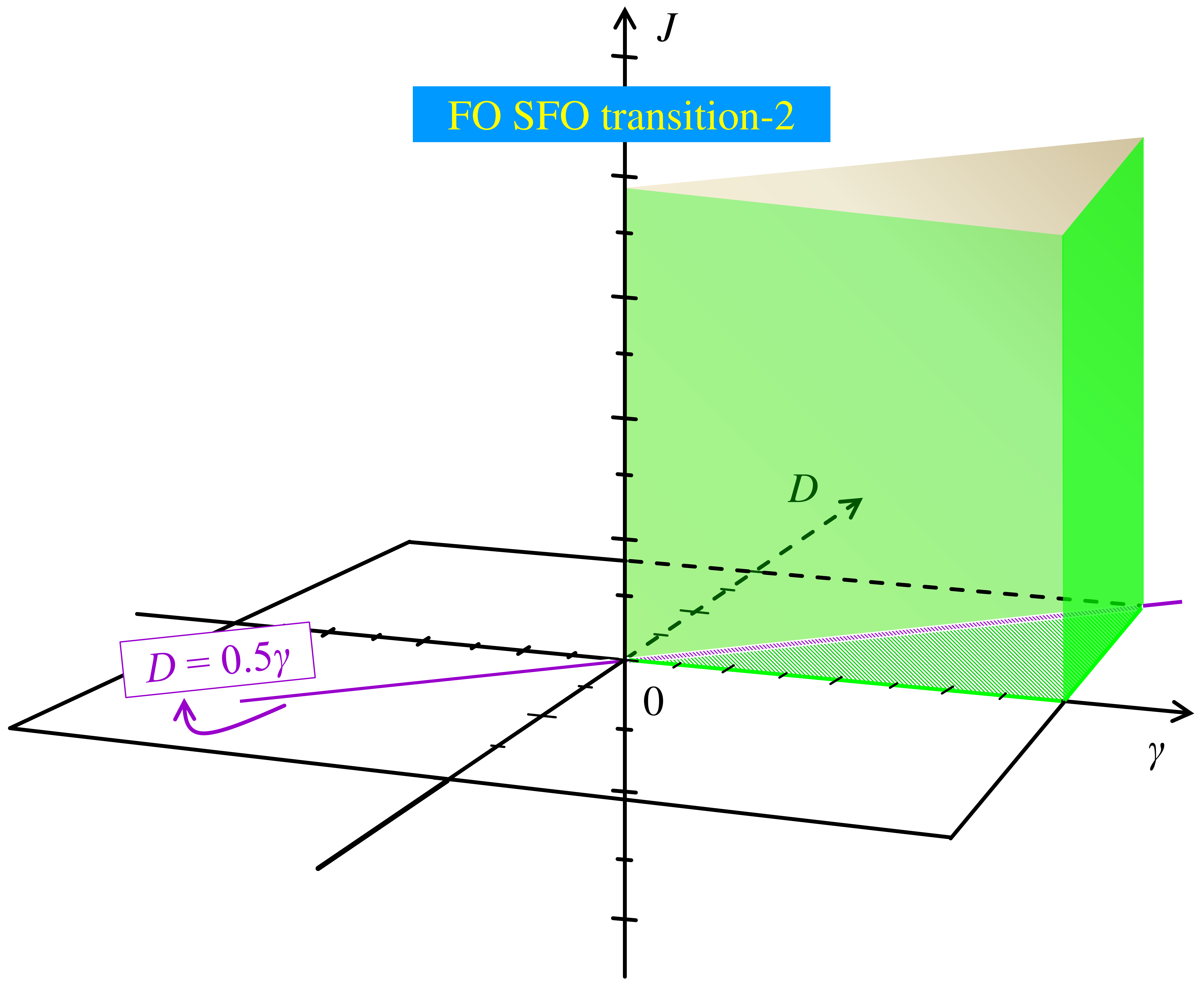}
\caption{
{Three{-}dimensional ($J, \gamma, D$) phase diagram of a collinear two{-}sublattice antiferromagnet.} When $0 \leq D \leq 0.5\gamma$, a FO SFO transition{-}2 occurs. It is pointed out that here $\gamma$ and $D$ can not be zero simultaneously.
}
\label{Li-Figure9}
\end{figure*}

\begin{figure*}[!ht]
\centering \includegraphics[width = 0.88\textwidth] {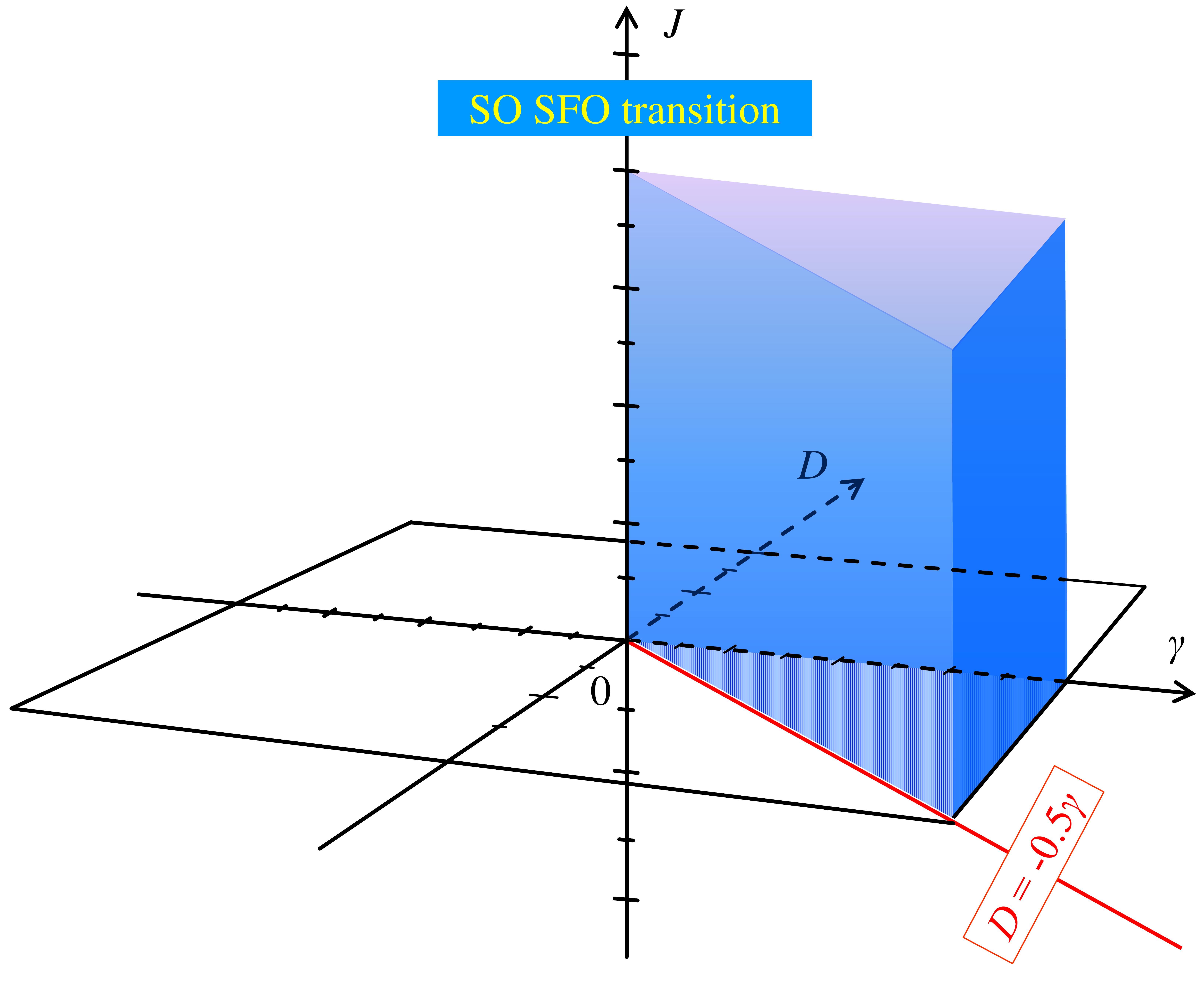}
\caption{
{Three{-}dimensional ($J, \gamma, D$) phase diagram of a collinear two{-}sublattice antiferromagnet.} When ${-}\frac{1}{2}\gamma < D < 0$, a SO SFO transition occurs.
}
\label{Li-Figure10}
\end{figure*}

\begin{figure*}[!ht]
\centering \includegraphics[width = 0.88\textwidth] {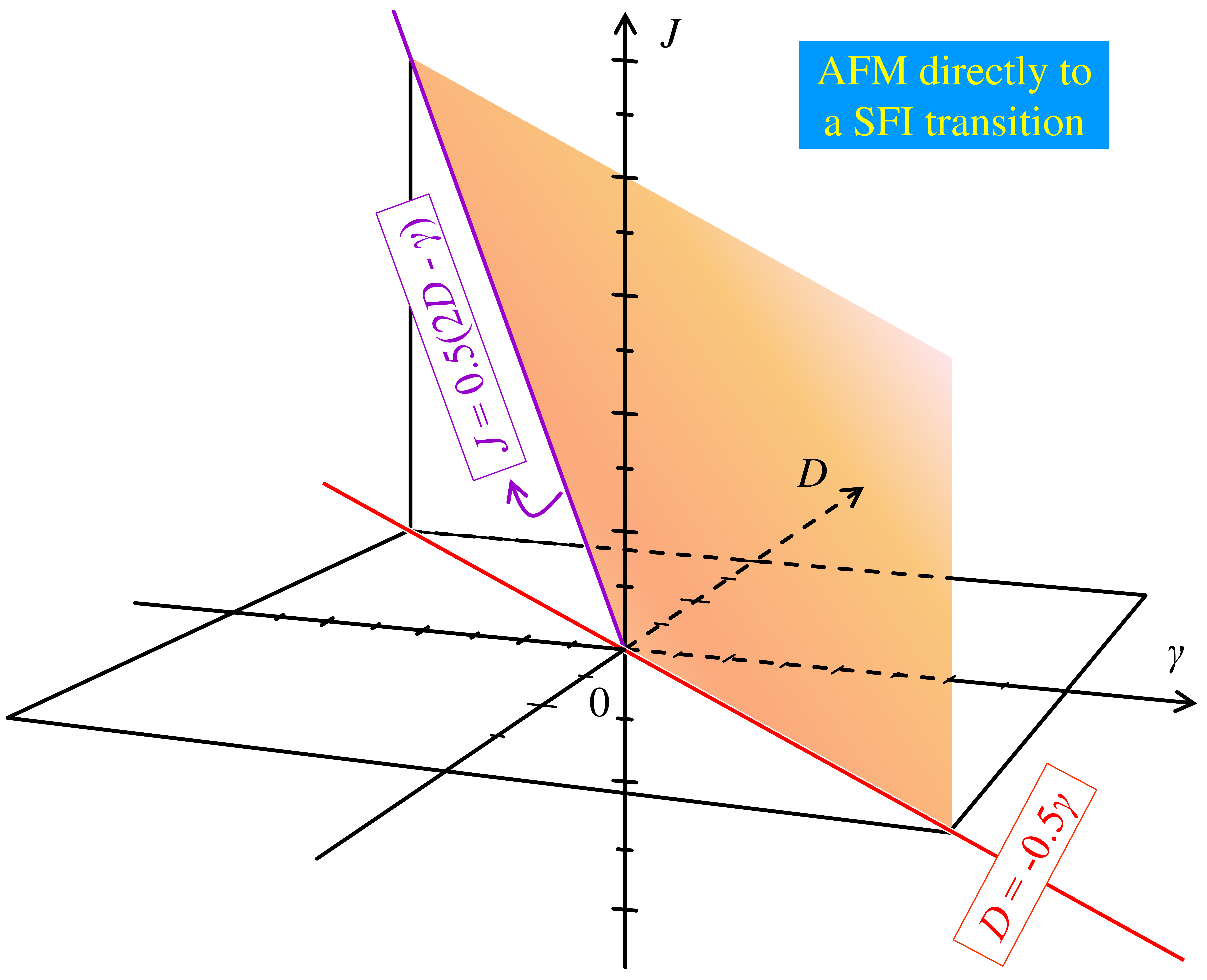}
\caption{
{Three{-}dimensional ($J, \gamma, D$) phase diagram of a collinear two{-}sublattice antiferromagnet.} When $D = {-}\frac{1}{2}\gamma$ and $J > {-}\gamma$, there is no SFO transition occurring at all. In this case, the antiferromagnet in question goes directly to a SFI transition from the AFM state, and the $xz${-}plane becomes an AFM easy plane.
}
\label{Li-Figure11}
\end{figure*}

\begin{figure*}[!ht]
\centering \includegraphics[width = 0.88\textwidth] {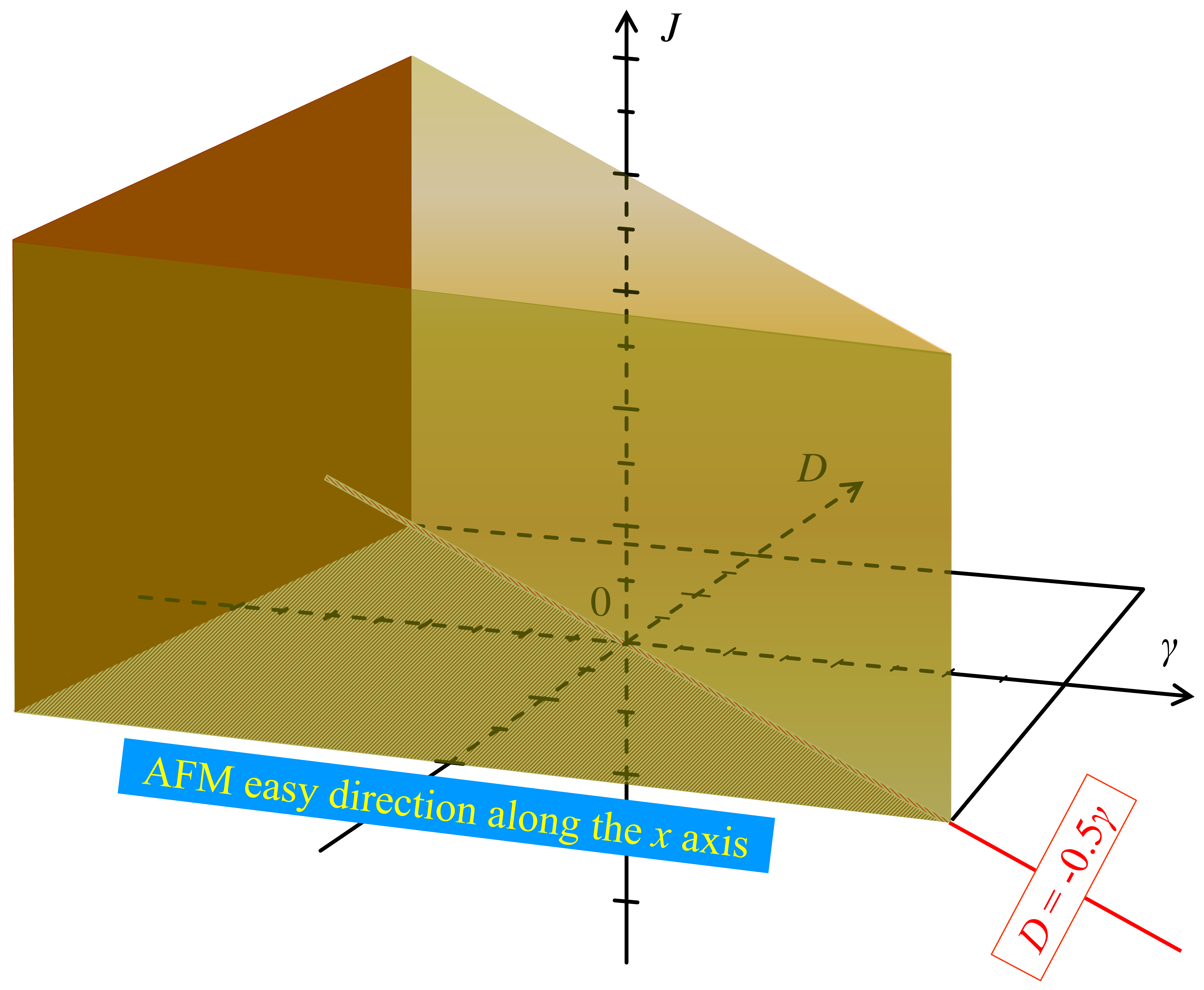}
\caption{
{Three{-}dimensional ($J, \gamma, D$) phase diagram of a collinear two{-}sublattice antiferromagnet.} When $D < {-}\frac{1}{2}\gamma$, the AFM easy axis changes automatically from the $z$ to the $x$ direction.
}
\label{Li-Figure12}
\end{figure*}

\begin{figure*}[!ht]
\centering \includegraphics[width = 0.88\textwidth] {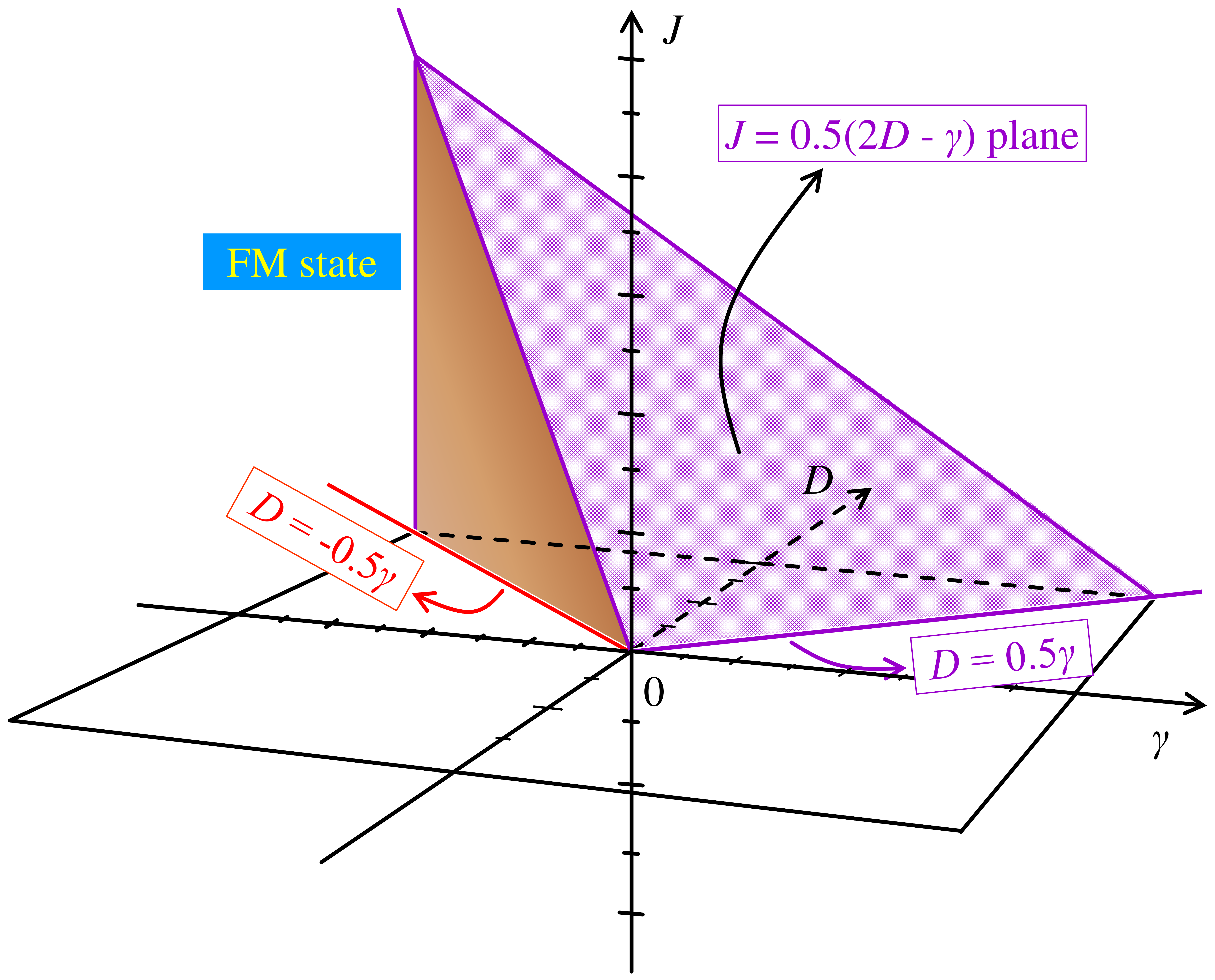}
\caption{
{Three{-}dimensional ($J, \gamma, D$) phase diagram of a collinear two{-}sublattice antiferromagnet.} When $D > \pm 0.5\gamma$ and $0 < J < \frac{1}{2}(2D {-} \gamma)$, the magnet hosts a FM state albeit with an AFM magnetic exchange.
}
\label{Li-Figure13}
\end{figure*}

\begin{figure*}[!ht]
\centering \includegraphics[width = 0.88\textwidth] {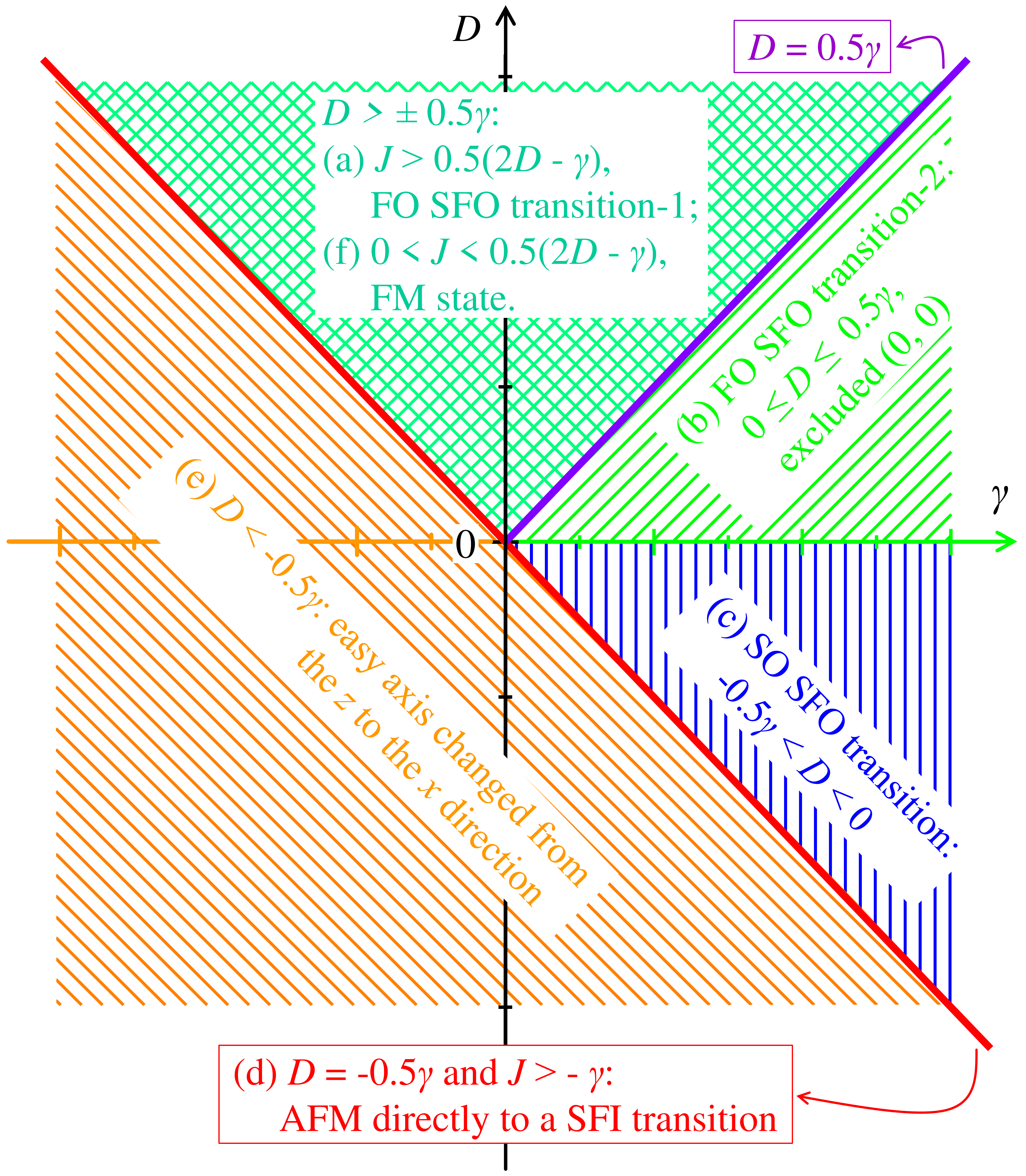}
\caption{
Two{-}dimensional ($\gamma, D$) phase diagram of a collinear two{-}sublattice antiferromagnet. The regimes (\textbf{a-f}) correspond to Figures~\ref{Li-Figure8}-\ref{Li-Figure13}, respectively.
}
\label{Li-Figure14}
\end{figure*}

\begin{figure*}[!ht]
\centering \includegraphics[width = 0.88\textwidth] {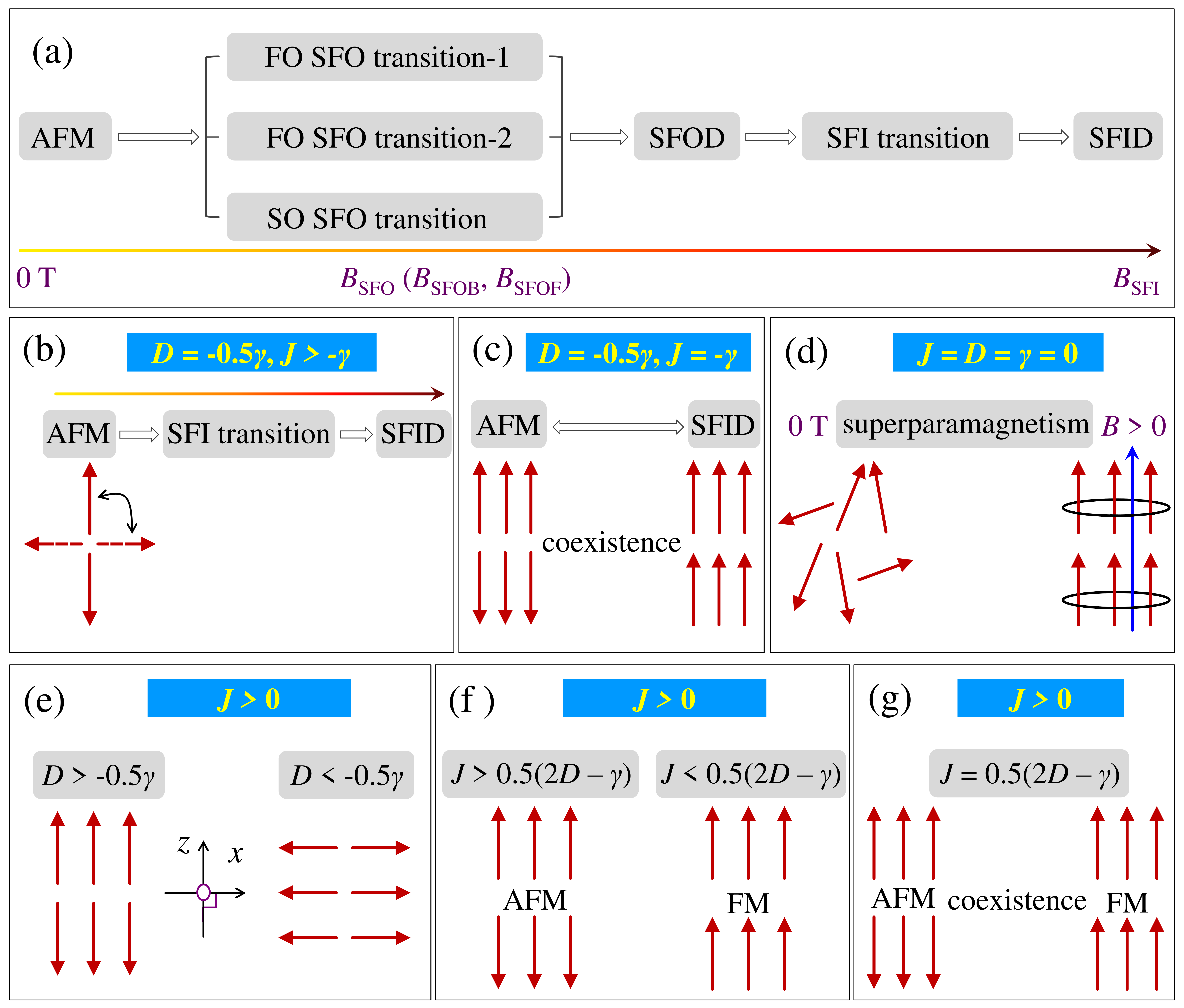}
\caption{
{Mathematically{-}allowed magnetic states of a collinear two{-}sublattice antiferromagnet as a function of $(J, \gamma, D, B)$.}
(\textbf{a}) As magnetic field $B$ increases from 0 to $B_{\textrm{SFO}}$ (for a FO SFO transition) or to $(B_{\textrm{SFOB}}, B_{\textrm{SFOF}})$ (for a SO SFO transition) and then to $B_{\textrm{SFI}}$, the antiferromagnet transfers from an AFM ground state to a FO SFO transition{-}1 or a FO SFO transition{-}2 or a SO SFO transition, and then to a SFOD state, from where a SFI transition occurs until all spins are flipped by applied magnetic field.
(\textbf{b}) When $D = {-}\frac{1}{2}\gamma$ and $J > {-}\gamma$, there exists an AFM easy plane.
(\textbf{c}) When $D = {-}\frac{1}{2}\gamma$ and $J = {-}\gamma$, the SFID state has the same energy level as that of the AFM ground state.
(\textbf{d}) Based on the above analysis of (\textbf{b}) and (\textbf{c}), one can deduce that when $J = D = \gamma = 0$, if magnetic field $B > 0$ applied, all spins will directly go to the SFID state and point to the applied field direction. This is the so{-}called superparamagnetism (Figure~\ref{Li-Figure3}(f)).
(\textbf{e}) When $J >0$ and $D > {-}\frac{1}{2}\gamma$, the AFM easy axis is along the $z$ direction; while $D < {-}\frac{1}{2}\gamma$, the $x$ axis becomes an AFM easy direction.
(\textbf{f}) When $J >0$ and $J > \frac{1}{2}(2D {-} \gamma)$, the magnet houses an AFM state; while when $J < \frac{1}{2}(2D {-} \gamma)$, the spins are ferromagnetically arranged.
(\textbf{g}) When $J >0$ and $J = \frac{1}{2}(2D {-} \gamma)$, it is reasonable to deduce that the AFM state coexists with the FM state.
}
\label{Li-Figure15}
\end{figure*}

In addition, for a SO SFO transition, ${-}\frac{1}{2}\gamma < D < 0$ (Figures~\ref{Li-Figure3}(b) and \ref{Li-Figure10}). It is pointed out that when ${-}\frac{1}{2}\gamma < D < \frac{1}{2}\gamma$, it is always true that $J > \frac{1}{2}(2D {-} \gamma)$. The difference between the two types of FO SFO transitions (1 and 2) in the context of $J$ is that for the FO SFO transition{-}1, $J > 0$ and $J > \frac{1}{2}(2D {-} \gamma)$; by contrast, for the FO SFO transition{-}2, $J$ can be any values larger than zero.
As shown in Figure~\ref{Li-Figure11}, when $D = {-}\frac{1}{2}\gamma$, $B_{\textrm{SFOB}} = B_{\textrm{SFOF}} = 0$ [equation~(\ref{SFOB})]. Therefore, the antiferromagnet directly enters a SFI transition (Figure~\ref{Li-Figure15}(b)). To further demonstrate this interesting magnetic phase transition, we calculate the relative free energies and the variations of the angles $\phi$ and $\beta$ (with the parameters $M_0 =$ 4 $\mu_\textrm{B}$, $J$ = 2 T/$\mu_\textrm{B}$, $D$ = {-}0.2 T/$\mu_\textrm{B}$, and $\gamma = 0.4$ T/$\mu_\textrm{B}$) as shown in Figures~\ref{Li-Figure7}(a) and \ref{Li-Figure7}(b). It is clear that this magnetic phase transition is theoretically favourable. It is more interesting that if $J = {-}\gamma$, $E_{\textrm{AFM}(z{-}\textrm{axis})}$ = $E_{\textrm{SFID}}$ [equation~(\ref{SFIE})], which implies that the AFM state can coexist with the SFID state (Figure~\ref{Li-Figure15}(c)). Based on the above discussion, it is reasonable to deduce that when $J = 0$ (a paramagnetic state) and $D = \gamma = 0$ (without any magnetic anisotropy), all paramagnetic spins will direct and be bounded to an applied-field direction when $B > 0$ (Figure~\ref{Li-Figure15}(d)). This is the so{-}called superparamagnetic state as schematically shown in Figure~\ref{Li-Figure3}(f).

When $D < {-}\frac{1}{2}\gamma$, $E_{\textrm{AFM}}$ is always larger than $E_{x{-}\textrm{axis}}$ [equation~(\ref{AFME})], indicating that the AFM easy axis will change from the $z$ to the $x$ axis (Figures~\ref{Li-Figure12} and \ref{Li-Figure15}(e)). Therefore, the AFM easy direction is determined by the competition between magnetic anisotropies, i.e., $\gamma$ and $D$.

\subsection{Estimation of the Magnetic Exchange Parameters $(J, \gamma, D)$ \cite{HFLi2016}}

As discussed above, when ${-}\frac{1}{2}\gamma < D < 0$ (Figure~\ref{Li-Figure3}(b)), a SO SFO transition occurs in an antiferromagnet. With the known exchange parameters $(J, \gamma, D)$, one can calculate the SFO $(B_{\textrm{SFOB}}, B_{\textrm{SFOF}})$ and SFI $(B_{\textrm{SFI}})$ fields, i.e.,
\setlength\arraycolsep{0.6pt} 
\begin{eqnarray}
\left \{
\begin{array} {lll}
\begin{split}
\label{SFOB}
B_{\textrm{SFOB}} & = M_0\sqrt{(2D {+} \gamma)(2J {+} 2D {+} \gamma)},                                                                                    \\
B_{\textrm{SFOF}} & = M_0(2J {-} 2D {+} \gamma)\sqrt{\frac{2D {+} \gamma}{2J {+} 2D {+} \gamma}},                                                             \\
B_{\textrm{SFI}} & = M_0 (2J {-} 2D {+} \gamma).
\end{split}
\end{array}
\right.
\end{eqnarray}
On the other hand, if the values of $B_{\textrm{SFOB}}$, $B_{\textrm{SFOF}}$ and $B_{\textrm{SFI}}$ are known, one can calculate the corresponding values of $(J, \gamma, D)$ according the following deduced equations from the above equation~(\ref{SFOB}), i.e.,
\setlength\arraycolsep{0.6pt} 
\begin{eqnarray}
\left \{
\begin{array} {lll}
\begin{split}
\label{DyJ}
D & = \frac{B_{\textrm{SFI}}}{4M_0} \left( \frac{B_{\textrm{SFOB}} {-} B_{\textrm{SFOF}}}{B_{\textrm{SFOF}}} \right),                                 \\
\gamma & = \frac{1}{M_0}\left[ \frac{B_{\textrm{SFOB}}B_{\textrm{SFOF}}}{B_{\textrm{SFI}}} {-} \frac{B_{\textrm{SFI}}(B_{\textrm{SFOB}} {-} B_{\textrm{SFOF}})}{2B_{\textrm{SFOF}}} \right],                                                                                                    \\
J & = \frac{1}{2M_0}\left[ B_{\textrm{SFI}}{-}\frac{B_{\textrm{SFOB}}B_{\textrm{SFOF}}}{B_{\textrm{SFI}}} {+} \frac{B_{\textrm{SFI}}(B_{\textrm{SFOB}} {-} B_{\textrm{SFOF}})}{B_{\textrm{SFOF}}} \right].                                                                                                                                             \end{split}
\end{array}
\right.
\end{eqnarray}

When $D \geq 0$, $D > {-}\frac{1}{2}\gamma$, and $J > \frac{1}{2}(2D {-} \gamma)$ (Figure~\ref{Li-Figure3}(d)), a FO SFO transition occurs, and
\setlength\arraycolsep{0.6pt} 
\begin{eqnarray}
\left \{
\begin{array} {ll}
\begin{split}
\label{SFO8}
B_{\textrm{SFO}} & = M_0\sqrt{(2D {+} \gamma)(2J {-} 2D {+} \gamma)},                               \\
B_{\textrm{SFI}} & = M_0 (2J {-} 2D {+} \gamma).
\end{split}
\end{array}
\right.
\end{eqnarray}
Although it is impossible to solve the above equation~(\ref{SFO8}) to extract the detailed values of $(J, \gamma, D)$, one can deduce that
\begin{align}
\label{DY}
2D {+} \gamma = \frac{B_{\textrm{SFO}}^2}{M_0B_{\textrm{SFI}}}.
\end{align}
Hence, one can calculate two special cases, i.e.,
\setlength\arraycolsep{0.6pt} 
\begin{eqnarray}
\left \{
\begin{array} {lll}
\begin{split}
\label{D0JY}
\textrm{if} \textcolor[rgb]{1.00,1.00,1.00}{a} D & = 0,                                                                                            \\
\textrm{then} \textcolor[rgb]{1.00,1.00,1.00}{a} \gamma & = \frac{B^2_{\textrm{SFO}}}{M_0 B_{\textrm{SFI}}}, \textcolor[rgb]{1.00,1.00,1.00}{a} \textrm{and}                                                                                                                                       \\
J & = \frac{B^2_{\textrm{SFI}} {-} B^2_{\textrm{SFO}}}{2M_0B_{\textrm{SFI}}};
\end{split}
\end{array}
\right.
\end{eqnarray}
and
\setlength\arraycolsep{0.6pt} 
\begin{eqnarray}
\left \{
\begin{array} {lll}
\begin{split}
\label{DJY0}
\textrm{if} \textcolor[rgb]{1.00,1.00,1.00}{a} \gamma & = 0,                                                                                            \\
\textrm{then} \textcolor[rgb]{1.00,1.00,1.00}{a} D & = \frac{B^2_{\textrm{SFO}}}{2M_0 B_{\textrm{SFI}}}, \textcolor[rgb]{1.00,1.00,1.00}{a} \textrm{and} \\
J & = \frac{B_{\textrm{SFI}}^2 {+} B_{\textrm{SFO}}^2}{2M_0B_{\textrm{SFI}}}.
\end{split}
\end{array}
\right.
\end{eqnarray}

Traditionally, through fitting the relevant $\textbf{Q}$ (momentum){-}$E$ (energy) spectra recorded usually by inelastic neutron scattering, one can extract the magnetic exchange parameters $(J, \gamma, D)$. Here, based on our model, one can first obtain the values of $B_{\textrm{SFO}} (B_{\textrm{SFOB}}, B_{\textrm{SFOF}})$ and $B_{\textrm{SFI}}$ for a suitable SFO and SFI compound, e.g., via magnetization measurements using a commercial physical property measurement system or a Quantum Design MPMS{-}7 SC quantum interference device magnetometer (Quantum Design, San Diego, USA). Then the values of $(J, \gamma, D)$ can be estimated according to equation~(\ref{DyJ}), (\ref{D0JY}) or (\ref{DJY0}) \cite{HFLi2016}.

\section* {Conclusion}
\begin{figure*}[!ht]
\centering \includegraphics[width = 0.88\textwidth] {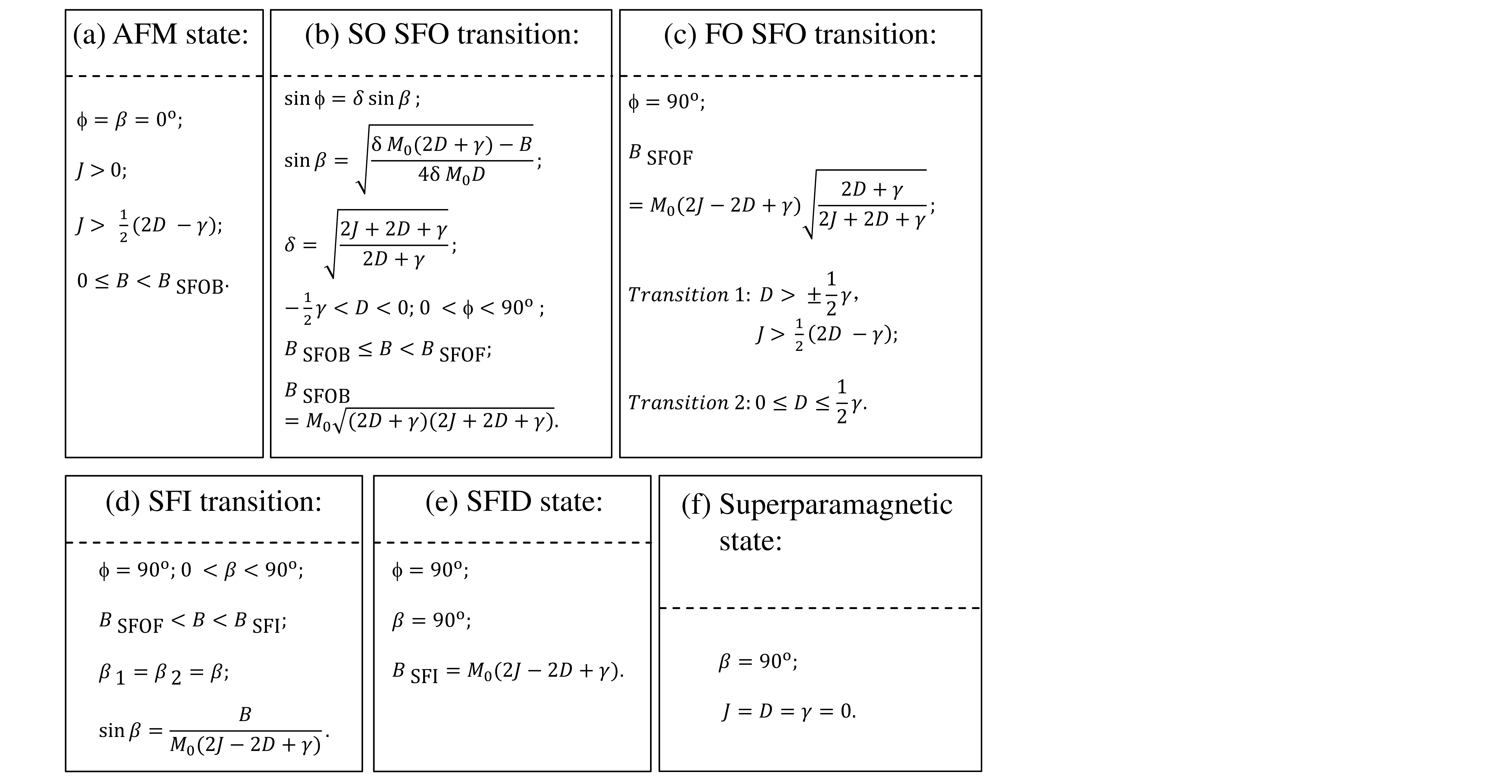}
\caption{
{The corresponding equilibrium magnetic phase conditions.}
(\textbf{a}) The normal AFM state. AFM = Antiferromagnetic.
(\textbf{b}) The SO SFO transition. SO SFO = Second order spin{-}flop transition.
(\textbf{c}) The FO SFO transition. FO SFO = First order spin{-}flop transition. Here exit two types of FO SFO transitions.
(\textbf{d}) The SFI transition. SFI = Spin{-}flip transition.
(\textbf{e}) The SFID state. SFID = Spin{-}flipped state.
(\textbf{f}) Derived superparamagnetic state.
}
\label{Li-Figure16}
\end{figure*}

To summarize, a consistent mean{-}field calculation of the SFO and SFI phase transitions has been performed for localized collinear antiferromagnets. In this study, we can unify all possible magnetic ground states as well as the related magnetic phase transitions within one model. Some special magnetic states are derived with a change in the strength of applied magnetic field: (1) A rotating FM{-}like state (that is finally ruled out); (2) A SO SFO transition; (3) A direct SFI transition from the AFM state without experiencing a SFO transition as usual; (4) An existence of the FM state; (5) A coexistence of the AFM and FM states even when the magnetic exchange is of AFM. The corresponding equilibrium phase conditions of the deduced mathematically allowable magnetic states are summarized in Figure~\ref{Li-Figure16}. Based on the quantitative changes of the ground{-}state free energies, the case (1) has been clearly ruled out, and the others indeed exist theoretically. This model calculation unifies the AFM state, FO and SO SFO transitions, SFOD state, SFI transition as well as the SFID state. Their respective phase boundary conditions are extracted and clearly listed. We find an alternative to the estimation of magnetic exchange parameters $(J, \gamma, D)$.

\section*{Acknowledgments}

Authors acknowledge the start{-}up research grants (No. SRG2016{-}00002{-}FST and No. SRG2016{-}00091{-}FST) at the University of Macau and financial support from the Science and Technology Development Fund, Macao SAR (Files No. 063/2016/A2, No. 064/2016/A2, No. 028/2017/A1, and No. 0051/2019/AFJ), and the Guangdong--Hong Kong--Macao Joint Laboratory for Neutron Scattering Science and Technology (Grant No. 2019B121205003), and thank X.Q.Y. for his help in collecting the references.

\newpage

\end{document}